\documentclass[aps,pra,reprint,groupedaddress,secnumarabic]{revtex4-2}

\usepackage{amsfonts,amssymb,amsmath}

\usepackage{mathtools}
\usepackage{enumitem}
\usepackage{times}

\usepackage[hiresbb]{graphicx}

\newcommand{\Star}[1]{#1\ensuremath{^\ast}\kern-\scriptspace}
\newcommand{\CStar}{\Star{\ensuremath{\mathrm{C}}}}

\DeclareMathOperator*{\argmin}{arg\,min}

\newcommand{\defeq}{\coloneqq}

\DeclareMathSymbol{\rstr}{\mathclose}{AMSa}{"16}

\newcommand{\HS}[1]{\mathcal{#1}}

\newcommand{\abs}[1]{\lvert #1 \rvert}
\newcommand{\dabs}[1]{\left\lvert #1 \right\rvert}

\newcommand{\norm}[1]{\lVert #1 \rVert}
\newcommand{\lnorm}[2]{\lVert #1 \rVert_{#2}}

\newcommand{\ev}[1]{\langle #1 \rangle}

\newcommand{\lev}[2]{\langle #1 \rangle_{\hspace{-1.0pt}#2}}
\newcommand{\dlev}[2]{\left\langle #1 \right\rangle_{\!#2}}

\newcommand{\inpr}[2]{\langle #1, #2 \rangle}

\newcommand{\linpr}[3]{\langle #1, #2 \rangle_{#3}}
\newcommand{\dlinpr}[3]{\left\langle #1,\, #2 \right\rangle_{\!#3}}

\newcommand{\stdv}[1]{\sigma(#1)}
\newcommand{\lstdv}[2]{\sigma_{\hspace{-1.0pt}#2}(#1)}
\newcommand{\dlstdv}[2]{\sigma_{\hspace{-1.0pt}#2}\left(#1\right)}

\newcommand{\id}{\mathrm{Id}}

\newcommand{\dom}[1]{\mathrm{dom}(#1)}

\newcommand{\ran}[1]{\mathrm{ran}\,#1}
\newcommand{\cran}[1]{\overline{\mathrm{ran}}\,#1}

\newcommand{\gr}[1]{\mathrm{gr}(#1)}

\newcommand{\cob}[2]{R_{#2}(#1)}
\newcommand{\qob}[2]{S_{\hspace{-0.5pt}#2}(\mathcal{#1})}

\newcommand{\spb}[1]{#1^{\hspace{-0.5pt}\ast}}
\newcommand{\spf}[1]{#1_{\hspace{-0.5pt}\ast}}
\newcommand{\pb}[2]{#1_{\!#2}^{\hspace{-0.5pt}\ast}}
\newcommand{\pf}[2]{#1_{\hspace{-1pt}#2\ast}}
\newcommand{\dpb}[2]{\pb{#1}{#2}}
\newcommand{\dpf}[2]{#1_{\!#2\ast}^{\phantom{\hspace{-0.5pt}\ast}}}

\newcommand\inv[1]{#1\raisebox{1.15ex}{$\hspace{0.5pt}\scriptscriptstyle-\!1$}}
\newcommand{\invpb}[2]{\inv{(\pb{#1}{#2})}}
\newcommand{\invpf}[2]{\inv{(\dpf{#1}{#2})}}

\newcommand{\err}[3]{\varepsilon_{\hspace{-0.5pt}#3}(#1;#2)}
\newcommand{\rerr}[3]{\tilde{\varepsilon}_{\hspace{-0.5pt}#3}(#1;#2)}

\newcommand{\dst}[3]{\eta_{#3}(#1;#2)}
\newcommand{\rdst}[3]{\tilde{\eta}_{#3}(#1;#2)}

\begin{document}

%
%

\title{A Universal Formulation of Uncertainty Relation for Error--Disturbance \\ and Local Representability of Quantum Observables}

%
%

\author{Jaeha Lee}
\email[]{lee@iis.u-tokyo.ac.jp}
\affiliation{Institute of Industrial Science, The University of Tokyo, Chiba 277-8574, Japan.}

%
%

\begin{abstract}
A universal formulation of the quantum uncertainty regarding quantum indeterminacy, quantum measurement, and its inevitable observer effect is presented with additional focus on the representability of quantum observables over a given state.  The operational tangibility of the framework assures that the resultant general relations admit natural operational interpretations and characterisations, and thereby perhaps most importantly, their experimental verifiability.  In view of the universal formulation, Heisenberg's original philosophy of the uncertainty principle, most typically exemplified in his famous gamma-ray microscope Gedankenexperiment, is revisited;  it is reformulated and restated as a refined no-go theorem, albeit perhaps in a weaker form than was originally intended.  The relations entail, in essence as corollaries to their special cases, several previously known relations, including most notably the standard Kennard--Robertson relation, the Arthurs--Kelly--Goodman relations for joint measurement as well as the Ozawa and the Watanabe--Sagawa--Ueda relations for error and error--disturbance.
\end{abstract}

\maketitle

%
%

\section{Introduction\label{sec:introduction}}

Almost a century has passed since the historical discovery of quantum theory, which has now become undoubtedly one of the central pillars supporting modern physics, and without it, current technologies and society would be totally unthinkable.  As is the current consensus, the laws of the microscopic world still defy our na{\"i}ve conception of nature and reality in many respects, such as the non-local correlation known as quantum entanglement and the fundamental trade-off relations collectively addressed as the quantum uncertainty, which are both counterintuitive and unfamiliar in the macroscopic realm of our everyday lives.

The uncertainty principle, originally advocated by Heisenberg \cite{Heisenberg_1927} in 1927, has served as a guiding light in the investigation of the quantum world by offering intuitive insights, which are oftentimes helpful, yet could sometimes be deceptive and misleading without careful application.  In this regard, the original presentation of the principle was soon followed by a mathematical formulation of Kennard \cite{Kennard_1927} revealing the lower bound $\hbar/2$ dictated by the Planck constant for the product of the standard deviations of position and momentum.  This was subsequently generalised to those of arbitrary observables $A$ and $B$ by Robertson \cite{Robertson_1929} with the familiar lower bound
\begin{equation}\label{ineq:urel_Kennard-Robertson}
\stdv{A}\,\stdv{B} \geq \abs{ \ev{[A, B]}} /2
\end{equation}
now being given by the expectation value of the commutator $[A, B] \defeq AB - BA$ of the observables over the quantum state concerned.  The mathematical clarity and simplicity of the Kennard--Robertson (KR) relation established its status as \textit{the} uncertainty relation, and has since secured its position as a standard material in textbooks.

On the other hand, the original exposition of Heisenberg---although his view on uncertainty (or \lq indeterminateness\rq\ \cite{Heisenberg_1930}) is known to be rather unclear given his nebulous description of his writings---did entertain the notion of measurement and its inevitable disruption of the system (observer effect) in his various examples including the famous gamma-ray microscope Gedankenexperiment, which the formalism of Kennard--Robertson cannot attend to.

This rather uncomfortable circumstances led to the emergence of alternative formulations of the principle involving measurement.  Arguably one of the most standard methods is the adoption of the indirect measurement scheme, which explicitly takes into consideration the system of an auxiliary meter device in measuring the original system of interest, thereby allowing for an intuitive conception of an otherwise vague and abstract mathematical definitions of error and disturbance.  One of the earlier results of this approach are the Arthurs--Kelly--Goodman (AKG) relations \cite{Arthurs_1965,Arthurs_1988} for error and statistical cost of measurement.  This was followed by the works of Ozawa, which presented the relations for error \cite{Ozawa_2004_01}, as well as that of error--disturbance \cite{Ozawa_2003};  his works have also seen refinements \cite{Branciard_2013} and modification \cite{Ozawa_2019}.  Another approach is the employment of the framework of estimation theory, upon which the works of Yuen--Lax \cite{Yuen_1973} and the more recent Watanabe--Sagawa--Ueda (WSU) \cite{Watanabe_2011, Watanabe_2011_06, Watanabe_2014} are grounded.  Apart form these, uncertainty relations have also been formulated from a measure-theoretic viewpoint \cite{Werner_2004, Miyadera_2008, Busch_2013}.

This paper presents a universal formulation of uncertainty relations yielding a family of novel inequalities that mark the trade-off relation between the measurement errors and its inevitable disturbance associated with a quantum measurement under the most general settings conceivable;  it is established upon arguably the simplest and most general framework of measurements of statistical nature without any additional assumptions or reference to  concrete measurement models whatsoever, \textit{i.e.}, the only objects concerned are the tangible measurement outcomes.  The virtue of this philosophy is that, on top of the obvious benefit of generality, the relations presented below are supported by operational interpretations and characterisations, which in turn guarantee their experimental verifiability;  these seemingly innocent and perhaps subtle remarks---while being of grave importance to empirical science---are not necessarily respected by some of the alternative formulations (including Ozawa's) that generally entertain concepts and objects that cannot be determined by the operational measurement outcomes alone, as have been criticised by several authors \cite{Werner_2004, Koshino_2005}.

Notably, the formulation of this paper yields several types of trade-off relations of different quality within a single framework, thereby seamlessly unifying the various forms of manifestation of uncertainty in quantum theory.  In fact, the uncertainty relations presented here entail, in essence as corollaries to their special cases, various known relations including the standard KR relation, the AKG relations for error and statistical cost, the Ozawa relations for error and error--disturbance, and the WSU relations for error and error--disturbance mentioned earlier.  Apart from the derivation of the KR relation, and also the outline leading to the AKG, Ozawa, and WSU relations, details on these topics shall be explicated in the subsequent manuscripts of the author within appropriate contexts.

This paper is organised as follows.  The first three sections following this introduction are devoted to the presentation of the universal framework.  Section~\ref{sec:process} concerns maps between state spaces, termed processes, for which the basic concepts and facts are given.  In Section~\ref{sec:pullback_and_pushforward}, processes are found to induce an adjoint pair of state-dependent maps, termed the pullback and pushforward.  Section~\ref{sec:partial-inverse} introduces another key mathematical concept of the universal formulation, namely the procedure of inverting a not-necessarily injective operator, termed the standard partial inverse.

Based on the tools and concepts prepared so far, the subsequent three sections introduce the main results of this paper.  In Section~\ref{sec:loss}, the concept of loss induced by a process, to which the error of a quantum measurement and the disturbance of a quantum process belong as special instances, is introduced;  this is followed by useful equivalence conditions on which a process becomes free from loss.  Section~\ref{sec:urel_error} is a review of some of the results presented in the previous work \cite{Lee_urel_2022_01} of the author, in which a family of uncertainty relations for error valid under the constraint of local joint-measurability (a weaker notion of the standard concept of joint measurability introduced by the author) are evoked.  The main results of this paper, the uncertainty relations for error--disturbance, are presented in Section~\ref{sec:urel_error-disturbance}.

The final two sections discuss some of the ramifications of the results obtained so far.  In Section~\ref{sec:uncertainty-principle}, an important connotation of the relations is presented, whereby Heisenberg's philosophy of the uncertainty principle is revisited;  in view of the universal framework, it is restated as a refined and rigorous no-go theorem, albeit perhaps in a weaker form than was originally intended.  The final Section~\ref{sec:reference_to_other_relations} comments on the previous studies, in which the new relations are found to entail, in essence as a corollary to their special cases, several notable relations mentioned earlier.

Part of the framework and the results presented in this paper have been reported in the earlier manuscripts \cite{Lee_2020_01_preprint,Lee_2020_01_Entropy} of the authors, to which the reader is referred as appropriate.  Further details on the physical ramifications of the universal formulation as well as their mathematical descriptions shall be reported in subsequent papers of the author.

\section{Process between State Spaces\label{sec:process}}

For the presentation of the framework, let $Z(\HS{H})$ denote the state space of a quantum system, which will be hereafter modelled as the convex set of all the density operators $\rho$ on a Hilbert space $\HS{H}$.  Its classical counterpart $W(\Omega)$ will be modelled as the convex set of all the probability distributions $p$ on a sample space $\Omega$.

\subsection{Process}

\begin{figure}
\includegraphics[hiresbb,clip,width=0.40\textwidth,keepaspectratio,pagebox=artbox]{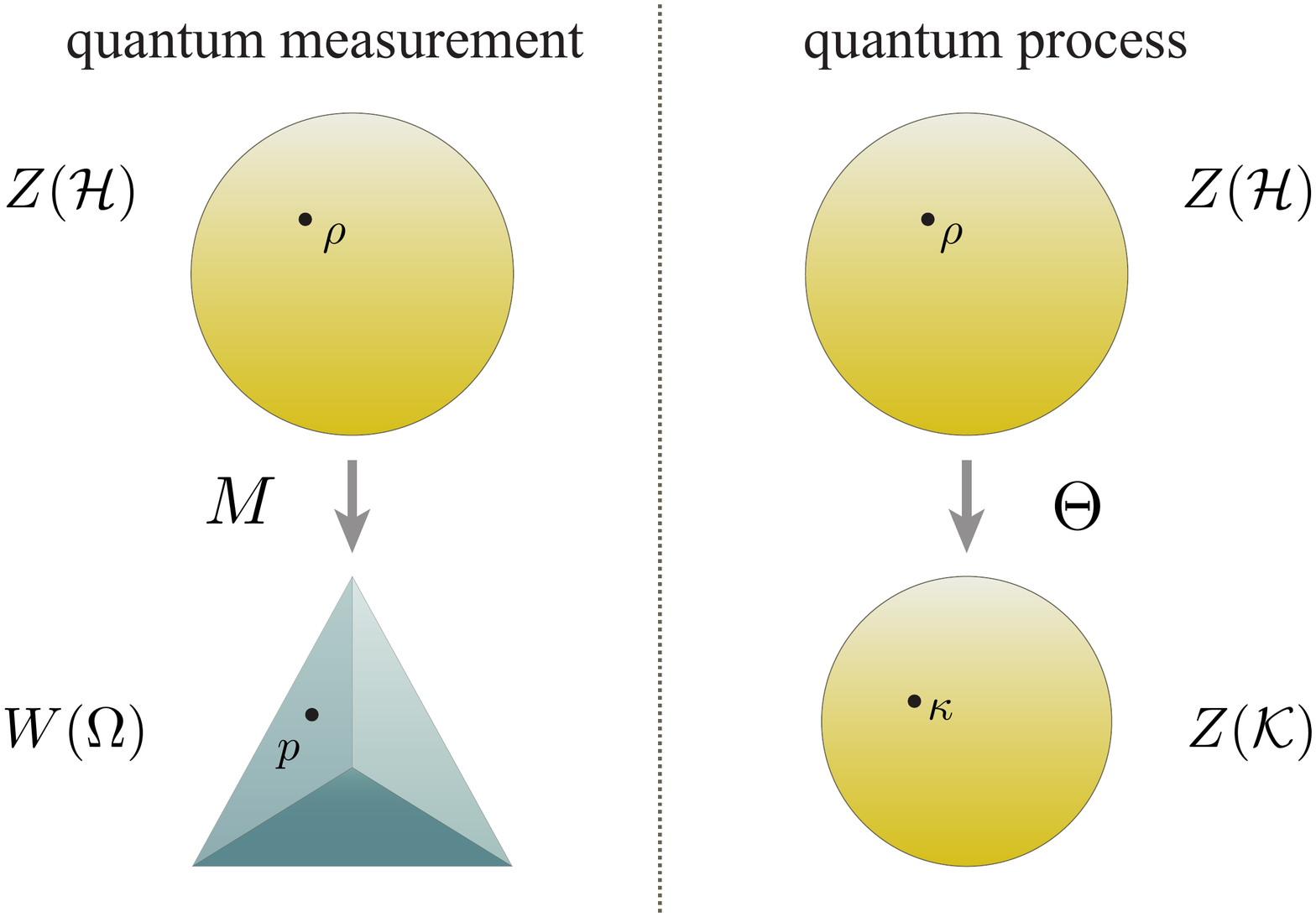}%
\caption{The basic premise of quantum measurements and quantum processes.  The space of quantum states (density operators) $Z(\HS{H})$ is depicted as a sphere, whereas the space of probability distributions $W(\Omega)$ is represented by a tetrahedron.  A quantum measurement $M : Z(\HS{H}) \to W(\Omega)$ can be regarded as a map that takes a given quantum state $\rho \in Z(\HS{H})$ to a probability distribution $p \in W(\Omega)$.  Similarly, a quantum process $M : Z(\HS{H}) \to Z(\mathcal{K})$ can be thought of as a map that takes a given quantum state $\rho \in Z(\HS{H})$ to another quantum state $\kappa \in Z(\mathcal{K})$.
}
\label{fig:processes}
\end{figure}

The primary objects of interest in this paper are the maps between state spaces (especially, quantum- and classical-state spaces in the current context) that maintains the consistency with the concept of probabilistic mixture of states, namely, affine maps between them.  In this paper, such maps shall be called \emph{processes} in generic terms, among which the following four types should be most relevant here:  those from quantum-state spaces to classical-state spaces (Q-C processes), those between classical-state spaces (C-C processes), those between quantum-state spaces (Q-Q processes), and those from classical-state spaces to quantum-state spaces (C-Q processes).  
In the current context of this paper, the first three types shall be given special attention.  A quantum-to-classical (Q-C) process $M : Z(\HS{H}) \to W(\Omega)$ shall be called a \emph{quantum measurement}, whereas a classical-to-classical (C-C) process $K : W(\Omega_{1}) \to W(\Omega_{2})$ is addressed as a \emph{classical measurement} (or depending on the context, a classical process).  A quantum-to-quantum (Q-Q) process $\Theta : Z(\HS{H}) \to Z(\mathcal{K})$ shall be simply called a \emph{quantum process} (see FIG. \ref{fig:processes}).

\subsubsection{Quantum Measurement}

The archetype of quantum measurements is the projection measurement.  More explicitly, let it be temporarily assumed for the sake of explanation that the measurement observable $\hat{M}$ is a non-degenerate Hermitian matrix acting on a finite $N$-dimensional space $\HS{H}$.  It is basic that its spectral decomposition $\hat{M} = \sum_{i=1}^{N} m_{i}\, |m_{i}\rangle\langle m_{i}|$ induces a natural affine map
\begin{equation}\label{def:projection-measurement}
M : \rho \mapsto (M\!\rho)(m_{i}) \defeq \mathrm{Tr}\bigl[ \vert m_{i}\rangle\langle m_{i} \vert \rho \bigr]
\end{equation}
that takes a density operator $\rho$ to the probability distribution $p(m_{i}) = (M\!\rho)(m_{i})$ defined on the spectrum $\Omega = \{ m_{1}, \dots, m_{N} \}$ of $\hat{M}$, which is a closed subset of the real line $\mathbb{R}$ consisting of all its eigenvalues.  The Born rule \cite{Born_1926} furnishes the distributions $M\!\rho$ with the standard interpretation as those describing the probability of finding each of the outcomes $m_{i} \in \Omega$ in the act of measurement.

Throughout this paper, the reader may---without missing much of the essence of the subject---simply think of the affine map $M$ as that associated with the familiar projection measurement \eqref{def:projection-measurement}.  Still, it is to be emphasised that the map $M$ is by no means restricted to that particular class, given that the sole constraint imposed here is the affineness $M(\lambda \rho_{1} + (1-\lambda)\rho_{2}) = \lambda M\!\rho_{1} + (1-\lambda) M\!\rho_{2}$, $\rho_{1}, \rho_{2} \in Z(\HS{H})$, $0 \leq \lambda \leq 1$ of the map.

\subsubsection{Quantum Process}

Analogously, the archetype of quantum processes within the context of this paper would be the `wave-function collapse' commonly associated with the `observer effect' induced by the act of measurement performed on the quantum system.  The familiar projection postulate \cite{Neumann_1932, Lueders_1951} dictates that the initial quantum state $\vert \psi\rangle$ over which the projection measurement \eqref{def:projection-measurement} is performed `collapses' to one of its eigenvectors $\vert m_{i}\rangle$ with probability $p(m_{i}) = \lvert \langle m_{i} | \psi \rangle \rvert^{2}$.  As is well-known, the projection postulate uniquely extends to mixed quantum states $\rho$, resulting in the affine map
\begin{equation}\label{def:collapse}
\Theta : \rho \mapsto \Theta\hspace{-0.05em}\rho := \sum_{i=1}^{N} \mathrm{Tr}\bigl[|m_{i}\rangle\langle m_{i}| \rho\bigr] \cdot |m_{i}\rangle\langle m_{i}|
\end{equation}
from the quantum state space $Z(\HS{H})$ to itself.  Throughout this paper, the reader may assume the quantum processes $\Theta$ to be those induced by the wave-function collapse \eqref{def:collapse} under the projection postulate, at least without much loss of essence;  in general, however, the map $\Theta$ are far from being restricted to such a class.

In this regard, it is to be reemphasised that, apart from the sole constraint of affineness $\Theta(\lambda \rho_{1} + (1-\lambda)\rho_{2}) = \lambda \Theta\hspace{-0.05em}\rho_{1} + (1-\lambda) \Theta\hspace{-0.05em}\rho_{2}$, $\rho_{1}, \rho_{2} \in Z(\HS{H})$, $0 \leq \lambda \leq 1$ of the map, which is indispensable for the self-consistent statistical interpretation of density operators, no other restriction is imposed to the notion of quantum process.  The map $\Theta$ treated here effectively describes the most general change of quantum states including, but not limited to, \textit{e.g.}, unitary evolution of closed quantum systems, non-unitary evolution of open quantum systems, quantum decoherence, observer effects (which is the main interest of this paper), quantum channels and gates;  it is to be noted that even the complete positivity of the (adjoint of the) quantum process, which is indispensable to many other formulations of the uncertainty relations, need not be assumed in the framework of the author.

\subsection{Adjoint of a Process}

An important observation is that a process, which has been introduced above as an affine map between state spaces, induces a natural map, termed its \emph{adjoint}, that is dual to the notion of the original process.  In order to avoid unnecessary repetition, arguments from now on focus primarily on the special instances of processes most relevant to the context of this paper, namely, quantum measurements and quantum processes.  It goes without saying that the same properties and facts described hereafter are also valid for any other processes, including classical measurements (classical processes) and classical-to-quantum (C-Q) processes as well, all of which can be demonstrated through a similar line of reasoning.

An important observation is that a quantum process $\Theta : Z(\HS{H}) \to Z(\mathcal{K})$ (or any process, as mentioned above) uniquely induces a map $\Theta^{\prime}$ between operator spaces.  This dual notion of a quantum process, termed its adjoint, is uniquely characterised by the relation
\begin{equation}\label{char:adjoint_quantum-process}
\dlev{ \Theta^{\prime} X }{\rho}
	= \dlev{ X }{(\Theta\hspace{-0.05em}\rho)}
\end{equation}
that is valid for all operators $X$ on $\mathcal{K}$ and quantum states $\rho$ on $\HS{H}$, in which the shorthand $\lev{X}{\rho} \defeq \mathrm{Tr}[X\rho]$ is defined for a pair of a Hilbert-space operator $X$ and a density operator $\rho$ on $\HS{H}$.  In this regard, the `wave-function collapse' \eqref{def:collapse} under the projection postulate provides a prime example, the adjoint of which can be confirmed to read
\begin{equation}\label{char:adjoint_collapse}
\Theta^{\prime} : X \mapsto \Theta^{\prime} X = \sum_{i=1}^{N} \mathrm{Tr}\bigl[|m_{i}\rangle\langle m_{i}| X\bigr] \cdot |m_{i}\rangle\langle m_{i}|,
\end{equation}
which indeed fulfils \eqref{char:adjoint_quantum-process}.  The projection postulate is convenient in that they admit concrete expressions both for the quantum process \eqref{def:collapse} and its adjoint \eqref{char:adjoint_quantum-process} in terms of the familiar operator-theoretic tools, allowing for the verification of the various claims of this paper by means of direct computation.

\section{Pullback and Pushforward of a Process\label{sec:pullback_and_pushforward}}

\begin{figure}
\includegraphics[hiresbb,clip,width=0.40\textwidth,keepaspectratio,pagebox=artbox]{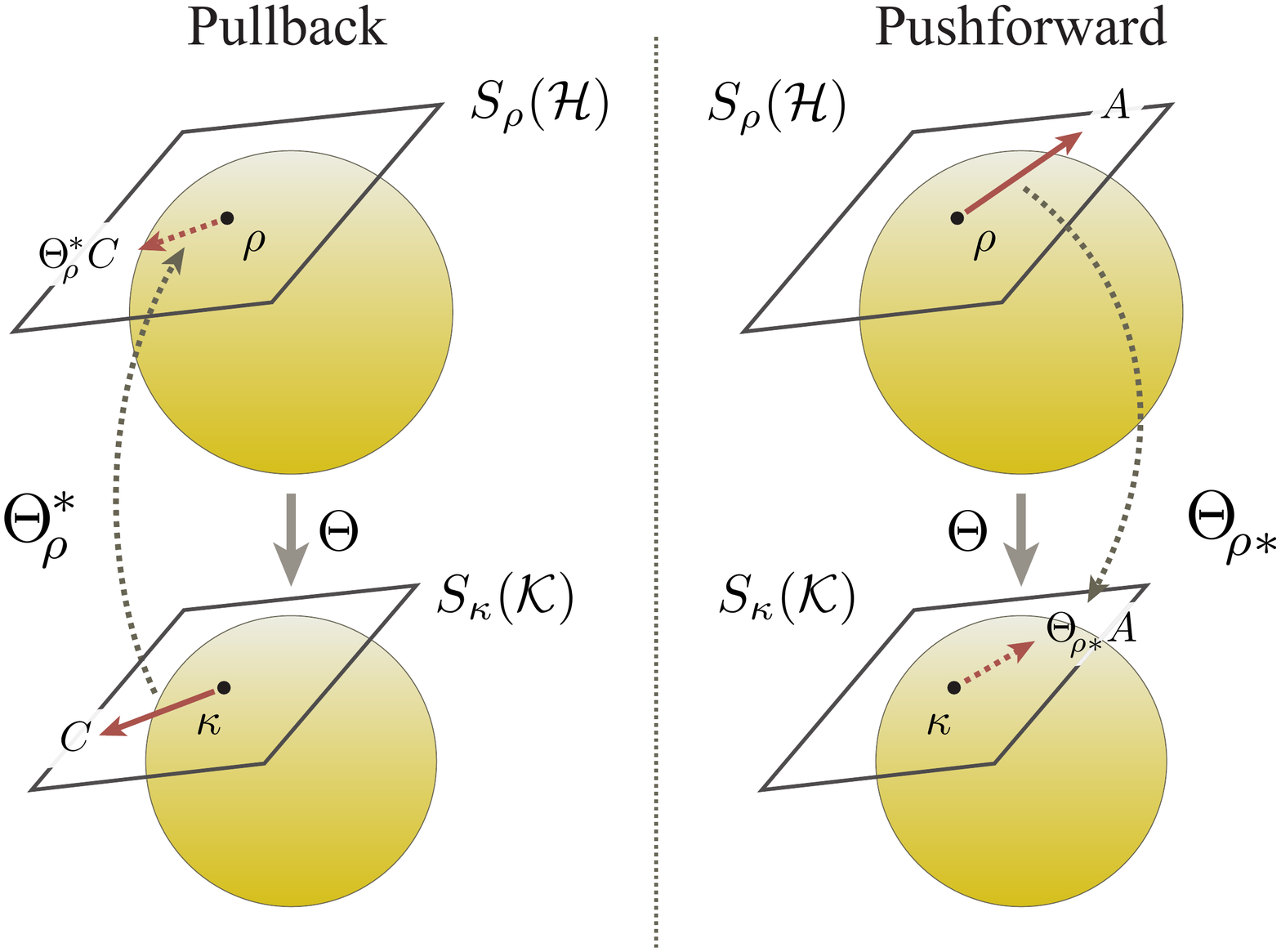}%
\caption{The pullback and the pushforward of the quantum process.  (Left) A quantum process $\Theta$ entails the pullback $\pb{\Theta}{\rho}$ from the space of localised observables $S_{\kappa}(\mathcal{K})$ to $S_{\rho}(\HS{H})$, each of which is attached to the respective points $\kappa = \Theta\hspace{-0.05em}\rho \in Z(\mathcal{K})$ and $\rho \in Z(\HS{H})$ of the corresponding state spaces.  (Right) Conversely, a quantum process $\Theta$ also entails the pushforward that maps in the opposite direction.  The pullback and the pushforward are dual to each other through the relation \eqref{char:pullback_pushforward}.}
\label{fig:pullback_pushforward}
\end{figure}

The key to the formulation of this paper is the observation that a process induces an adjoint pair of \emph{local} (\textit{i.e.}, state-dependent) maps between the localised observable spaces introduced below.  The pair, termed the pullback and pushforward of a process, serves as an important constituent of the framework (see FIG.~\ref{fig:pullback_pushforward}).

\subsection{The Space of Localised Observables}

The space of quantum observables will be modelled by the linear space $S(\HS{H})$ of all the self-adjoint operators on a Hilbert space $\HS{H}$.  Here, each quantum state $\rho \in Z(\HS{H})$ defines a seminorm $\lnorm{ A }{\rho} \defeq \sqrt{\lev{ A^{\dagger} A }{\rho}}$, $A \in S(\HS{H})$ on the space, thereby inducing a natural equivalence relation $A \sim_{\rho} B \iff \lnorm{ A - B }{\rho} = 0$ on it.  This allows for the classification of all the quantum observables into their equivalence classes $[A]_{\rho} \defeq \{ B \in S(\HS{H}) : A \sim_{\rho} B \}$, which collectively constitute a quotient space $S(\HS{H})/{\sim_{\rho}}$, the unique completion
\begin{equation}\label{def:loc_quantum_observables}
\qob{H}{\rho}
	\defeq \overline{ S(\HS{H})/{\sim_{\rho}} }
\end{equation}
of which shall be addressed in this paper as the space of (localised) quantum observables over $\rho$.  In a parallel manner, a probability distribution $p \in W(\Omega)$ induces a seminorm $\lVert f \rVert_{p} \defeq \sqrt{\lev{ f^{\dagger}f }{p}}$ on the linear space $R(\Omega)$ of all the real functions $f$ defined on the sample space $\Omega$.  The identification $f \sim_{p} g \iff \lVert f - g \rVert_{p} = 0$ results in the classification of the real functions into their equivalence classes $[f]_{p} \defeq \{ g \in R(\Omega) : f \sim_{p} g \}$, which collectively make up the quotient space $R(\Omega)/{\sim_{p}}$, further leading to its unique completion
\begin{equation}\label{def:loc_classical_observables}
\cob{\Omega}{p}
	\defeq \overline{ R(\Omega)/{\sim_{p}} }
\end{equation}
termed the space of (localised) classical observables over $p$ in this paper.  Here, the adjoint $A^{\dagger}$ of a Hilbert-space operator and the complex conjugate $f^{\dagger}$ of a complex function are introduced to expose the structure of the seminorms so that they respectively admit obvious extensions beyond self-adjoint operators and real functions.  As commonly practiced, with a slight abuse of notation, the equivalence classes will be denoted by one of their representatives hereafter.

\subsection{Pullback of a Process}

Given a quantum process $\Theta : Z(\HS{H}) \to Z(\HS{K})$, a crucial observation regarding its adjoint is the validity of the inequality
\begin{equation}\label{ineq:non-expansiveness_adjoint}
\lnorm{ C }{\Theta\hspace{-0.05em}\rho} \geq \lnorm{ \Theta^{\prime} C }{\rho}
\end{equation}
for any quantum state $\rho$ on $\HS{H}$ and self-adjoint operator $C$ on $\HS{K}$ (this is also valid for normal operators as well).   A simple way to understand this would be by means of the Kadison--Schwarz inequality \cite{Kadison_1952}, which is, in a certain sense, a generalisation of the prestigious Cauchy--Schwarz inequality to \CStar-algebras.  Indeed, its straightforward application to the adjoint $\Theta^{\prime}$ yields the evaluation $\Theta^{\prime}(N^{\dagger}N) \geq (\Theta^{\prime}N)^{\dagger}(\Theta^{\prime}N)$ valid for any normal operator $N$ on $\mathcal{K}$, which, combined with the characterisation \eqref{char:adjoint_quantum-process} of the adjoint, leads to the desired result \eqref{ineq:non-expansiveness_adjoint} as an immediate corollary.

A direct connotation of the inequality \eqref{ineq:non-expansiveness_adjoint} is the implication $C \sim_{\Theta\hspace{-0.05em}\rho} D \implies \Theta^{\prime}C \sim_{\rho} \Theta^{\prime}D$ for each quantum state $\rho \in Z(\HS{H})$.  This allows for the adjoint $\Theta^{\prime}$, which is a map between Hilbert-space operators, to be passed to the map between their quotient spaces over the quantum states concerned;  the unique continuous extension
\begin{equation}\label{def:pullback}
\dpb{\Theta}{\rho} : \qob{K}{(\Theta\hspace{-0.05em}\rho)} \to \qob{H}{\rho}
\end{equation}
of the resultant map shall be called the \emph{pullback of the quantum process} $\Theta$ over the quantum state $\rho$.  Among the remarkable properties of the pullback, the non-expansiveness $\lnorm{ C }{\Theta\hspace{-0.05em}\rho} \geq \lnorm{ \pb{\Theta}{\rho} C }{\rho}$ and the preservation $\lev{ C }{\Theta\hspace{-0.05em}\rho} = \lev{ \dpb{\Theta}{\rho} C }{\rho}$ of the expectation value, both of which may be confirmed straightforwardly by construction, are worthy of special note.

\subsection{Pushforward of a Process}

Dual to the notion of the pullback is the \textit{pushforward} of a quantum process.  For this, note that the norm on $\qob{H}{\rho}$ admits a unique inner product $\linpr{ A }{ B }{\rho} \defeq \lev{ \{A, B\} }{\rho}/2$ dictated by the anti-commutator $\{A,B\} \defeq AB + BA$ that reproduces the original norm $\lnorm{ A }{\rho}^{2} = \linpr{ A }{ A }{\rho}$.  The pushforward 
\begin{equation}\label{def:pushforward}
\dpf{\Theta}{\rho} : \qob{H}{\rho} \to \qob{K}{(\Theta\hspace{-0.05em}\rho)}
\end{equation}
is then introduced as the adjoint map of the pullback \eqref{def:pullback} with respect to the above inner products;  in other words, the pushforward is characterised as the unique map that satisfies the relation
\begin{equation}\label{char:pullback_pushforward}
\dlinpr{ A }{ \dpb{\Theta}{\rho}C }{\rho}
	= \dlinpr{ \dpf{\Theta}{\rho} A }{ C }{\Theta\hspace{-0.05em}\rho}
\end{equation}
for all $A \in \qob{H}{\rho}$ and $C \in \qob{K}{(\Theta\hspace{-0.05em}\rho)}$.  Owing to the non-expansiveness of the pullback, its adjoint, \textit{i.e.}, the pushforward, is also non-expansive $\lnorm{ A }{\rho} \geq \lnorm{ \pf{M}{\rho} A }{\Theta\hspace{-0.05em}\rho}$.  As with the pullback, the pushforward also preserves the expectation value $\lev{ A }{\rho} = \lev{ \pf{\Theta}{\rho} A }{\Theta\hspace{-0.05em}\rho}$, as can be readily confirmed by choosing the identity operator $C = \id$ in \eqref{char:pullback_pushforward}.

\subsection{Composition Laws\label{sec:composition_pullback_and_pushforward}}

By construction, composition of processes gives rise to another process.  The composition laws of the pullback and pushforward regarding process composition are crucial to the framework of the author.

For the sake of simplicity of arguments and due to the relevance to this paper, the composition laws shall be spelled out for the composition of a quantum measurement $M : Z(\HS{H}) \to W(\Omega)$ and a quantum process $\Theta : Z(\HS{H}) \to Z(\HS{K})$, which results in a new composite quantum measurement $M \circ \Theta: Z(\HS{H}) \to W(\Omega)$;  it goes without saying that, as always, the composition laws are valid for any types of process compositions in an obvious manner.  Under the above settings, the pullback
\begin{equation}\label{eq:composition_pullback}
\dpb{(M \circ \Theta)}{\rho}
	= \dpb{\Theta}{\rho} \circ \dpb{M}{(\Theta\hspace{-0.05em}\rho)}
\end{equation}
and the pushforward
\begin{equation}\label{eq:composition_pushforward}
\dpf{(M \circ \Theta)}{\rho}
	= \dpf{M}{(\Theta\hspace{-0.05em}\rho)} \circ \dpf{\Theta}{\rho}
\end{equation}
of the composite process $M \circ \Theta$ are respectively given by the compositions of those of the individual processes;  the composition law \eqref{eq:composition_pullback} for the pullback is a direct consequence of the composition law $(M \circ \Theta)^{\prime} = \Theta^{\prime} \circ M^{\prime}$ of the adjoint, which in turn follows directly from its characterisation \eqref{char:adjoint_quantum-process}, whereas that \eqref{eq:composition_pushforward} for the pushforward is a direct consequence of the property $(TS)^{*} \mapsto S^{*}T^{*}$ of the involution defined on the \CStar-algebra of bounded Hilbert-space operators regarding operator compositions.

It is to be noted that, through induction, the composition laws of both the pullback \eqref{eq:composition_pullback} and the pushforward \eqref{eq:composition_pushforward} admit generalisations to the composition of any number of arbitrary processes in an obvious manner.

\section{Standard Partial Inverse\label{sec:partial-inverse}}

Other important tools in the framework of the author are the `inverses' defined for the pullback and the pushforward induced by the quantum measurement, which shall be termed their (standard) partial inverses.  For this, a general method \cite{Lee_urel_2022_01} of inverting a not-necessarily injective closed operator is reviewed below.

\subsection{Generalised Inverses}

The aim of this passage is to construct an `inverse' of a not necessarily invertible operator.  Historically, the endeavour of inverting a generally non-injective map appeared in print as early as in 1903 in the work of Fredholm \cite{Fredholm_1903}, in which he introduced, in his original terminology, a particular `pseudo-inverse' of a certain integral operator, and also perhaps implicitly in the works of Hilbert \cite{Hilbert_1904_01, Hilbert_1904_02}, in which he alluded to such an idea in 1904 in studying generalised Green functions.  Arguably, one of the most renowned constructions in the community of physics would be the `general reciprocal' of finite-dimensional matrices, originally presented by Moore \cite{Moore_1920} in 1920, and later rediscovered by several individuals including Bjerhammar \cite{Bjerhammar_1951_01, Bjerhammar_1951_02, Bjerhammar_1958}, who recognised its role in the solution of linear systems.  Bjerhammar's results has seen further refinement and extension by Penrose \cite{Penrose_1955} in 1955, in which he specifically identified the characterisation of the reciprocal by the familiar four conditions.  Since then, the topic has received much attention and many contributions have been made.  Given the diverse terminologies for the various definitions of the `inverse', the term `generalised inverse' seems to be widely accepted as a generic word for collectively addressing the different ways of construction.

\subsection{Standard Partial Inverse of a Closed Operator}

For the purpose of this paper, the author introduces below the notion of \emph{(standard) partial inverse} defined for closed linear operators, which may serve as another contribution to the study of various constructions of the generalised inverses;  in this paper, partly in order to avoid confusion with the numerous alternative constructions, the author introduces a different terminology than the preexisting terms, some of which have been reviewed above.

Recall that a linear operator $A: U \supset \dom{A} \to V$ between normed spaces $U$ and $V$ is said to be closed, if its graph
\begin{equation}\label{def:graph_of_operator}
\gr{A} \defeq \{ (x, Ax) : x \in \dom{A} \}
\end{equation}
is a topologically closed subspace of $U \oplus V$ with respect to the product topology, or equivalently, if the domain $\dom{A}$ is complete with respect to the graph norm $\lnorm{x}{A} \defeq \norm{x} + \norm{Ax}$.  Closed operators constitute a decently well-behaved class of linear maps, while encompassing a fairly wide range of important operators encountered in many practical situations;  important subclasses are those of bounded operators as well as (not-necessarily bounded) self-adjoint operators.

Let $A : \HS{H} \supset \dom{A} \to \HS{K}$ be a closed operator between Hilbert spaces $\HS{H}$ and $\HS{K}$.
Given the fact that the kernel of a closed operator is topologically closed, one readily obtains the decomposition
\begin{equation}
\dom{A}
	= \ker{A} \oplus D_{\!A}
\end{equation}
of its domain with $D_{\!A} \defeq \ker{A}^{\perp} \cap \dom{A}$.  By construction, the restriction $A \rstr_{D_{\!A}}$ of the operator $A$ to the subspace $D_{\!A}$ is trivially injective, and is also itself a closed operator;  indeed, the graph $\gr{A \rstr_{D_{\!A}}} = \gr{A} \cap \{ (x,y) \in \HS{H} \oplus \HS{K} : x \in \ker{A}^{\perp} \}$ of the restriction is topologically closed as the intersection of two closed subspaces.

Under the above settings, the author introduces the \emph{(standard) partial inverse}
\begin{equation}\label{def:partial-inverse}
\inv{A}
	\defeq \inv{\left( A \rstr_{D_{\!A}} \right)}
\end{equation}
of a closed operator $A$ as the inverse of the restriction $A \rstr_{D_{\!A}}$, which is by construction an injective operator.  Conveniently, the partial inverse of a closed operator is itself closed, as one may readily see from the fact that the graph $\gr{\inv{A}} = U^{-1} \bigl( \gr{A \rstr_{D_{\!A}}} \bigr)$ of the partial inverse and that of the restriction are characterised by one another through the unitary operator $U : \HS{H} \oplus \HS{K} \to \HS{K} \oplus \HS{H},\, (x,y) \mapsto (y,x)$.

If the operator $A$ is invertible (\textit{i.e.}, injective), note that by construction, the partial inverse $\inv{A}$ coincides with the inverse operator of $A$ in the usual sense.  In this regard, the partial inverse rightfully serves as a generalisation of the standard notion of the inverse defined for invertible linear maps.

The following properties regarding a closed operator $A$ and its partial inverse $\inv{A}$ are straightforward by construction:
\begin{align}
\inv{A}A &= P_{D_{\!A}}, \label{char:partial-inverse_01} \\
A\inv{A} &= \id_{\ran{A}}. \label{char:partial-inverse_02}
\end{align}
Here, $P_{D_{\!A}}: \ker{A} \oplus D_{\!A} \ni (x_{1},x_{2}) \mapsto x_{2} \in D_{\!A}$ is the projection onto the summand $D_{\!A}$ associated with the direct sum decomposition $\dom{A} = \ker{A} \oplus D_{\!A}$ of the domain, and $\id_{\ran{A}}$ is the identity operator defined on $\ran{A}$.  Specifically, these immediately entail the equalities
\begin{align}
A\inv{A}A &= A, \\
\inv{A}A\inv{A} &= \inv{A},
\end{align}
which are two of the familiar characterising conditions commonly adopted by many of the constructions of the generalised inverse.

\subsection{Minimality}

The partial inverse also enjoys the familiar property regarding minimality that is widely found in common among many constructions of the generalised inverse.

Let $y \in \dom{\inv{A}} = \ran{A}$ be an element pertaining to the domain of the partial inverse of a closed operator $A$.  Then, among the elements $x \in \dom{A}$ for which $Ax = y$ hold, the minimum of the norm is uniquely attained by the partial inverse, \textit{i.e.},
\begin{align}\label{char:minimality}
\norm{x} \geq \norm{\inv{A}y}
\end{align}
and $\norm{x} = \norm{\inv{A}y} \iff x = \inv{A}y$; this is a straightforward consequence of the property~\eqref{char:partial-inverse_01}, which entails that any element $x \in \dom{A}$ of the domain admits a unique decomposition $x = x_{0} + x_{A}$ into the sum of the orthogonal elements $x_{0} \defeq x - \inv{A}Ax \in \ker{A}$ and $x_{A} \defeq \inv{A}Ax \in D_{\!A}$, thereby revealing
\begin{equation}\label{eq:decomposition_minimality}
\norm{x}^{2}
	= \norm{x - \inv{A}Ax}^{2} + \norm{\inv{A}Ax}^{2} \geq \norm{\inv{A}Ax}^{2}
\end{equation}
as desired.  In other words, the partial inverse of $A$ maps an element $y \in \dom{\inv{A}}$ to the unique argument of the minimum
\begin{equation}\label{char:minimality_argmin}
\argmin_{Ax=y} \norm{x}
	= \{ \inv{A}y \}
\end{equation}
of the norm of $\HS{H}$ restricted to the preimage of $y$ under $A$.

\subsection{Compositions\label{sec:partial-inverse:composition}}

It is useful to investigate the behaviour of the partial inverse regarding composition of operators.  By construction, the composition $AB$ of two invertible maps $A$ and $B$ is itself invertible, the inverse of which reads
\begin{equation}\label{eq:inverse_composition}
\inv{(AB)} = \inv{B} \inv{A}
\end{equation}
as is familiar.  However, this relation may no longer hold when the maps concerned are not necessarily injective.

Let $A : \HS{H}_{2} \supset \dom{A} \to \HS{H}_{3}$ and $B : \HS{H}_{1} \supset \dom{B} \to \HS{H}_{2}$ be closed operators between Hilbert spaces, and assume moreover that the composite operator $AB$ is closed.  Note here that the composition of closed operators are in general not necessarily closed;  a sufficient condition (that is relevant to this paper) for $AB$ to be closed is when the operator $B$ is bounded, as one may readily confirm using rudimentary techniques of functional analysis.

From a simple set-theoretic argument, note that the domains of the operators $\inv{(AB)}$ and $\inv{B} \inv{A}$ satisfies the relation
\begin{equation}
\dom{\inv{(AB)}} \supset \dom{\inv{B} \inv{A}}
\end{equation}
by construction.  For every element $z \in \dom{\inv{B} \inv{A}}$, one then finds the evaluation
\begin{equation}\label{eq:evaluation_inverse_composition}
\norm{\inv{B} \inv{A} z} \geq \norm{\inv{(AB)}z}
\end{equation}
from the minimality property of the partial inverse (or more precisely, the evaluation
\begin{equation}\label{eq:decomposition_composition}
\norm{\inv{B} \inv{A} z}^{2}
	= \norm{\inv{B} \inv{A} z - \inv{(AB)}z}^{2} + \norm{\inv{(AB)}z}^{2}
\end{equation}
from \eqref{eq:decomposition_minimality} regarding the partial inverse of $AB$), in consideration of the fact that both the elements $\inv{B} \inv{A} z$ and $\inv{(AB)}z$ of $\HS{H}_{1}$ are mapped to the element $z$ by the operator $AB$.  The uniqueness of the argument of the minimum (or more directly, the equality \eqref{eq:decomposition_composition}) thus reveals the equivalence of the conditions
\begin{equation}\label{char:partial-inverse_clean_1}
\inv{(AB)}z = \inv{B} \inv{A}z
	\iff \norm{\inv{(AB)}z} = \norm{\inv{B} \inv{A} z},
\end{equation}
which is, of course, a weaker form of the relation \eqref{eq:inverse_composition}.

It turns out to be useful to characterise the above equivalence in terms of the elements of $\HS{H}_{2}$, which can be pinned down through a parallel argument.  In this regard, observing that the elements $B\inv{(AB)}z$ and $\inv{A}z$ defined for $z \in \dom{\inv{(AB)}}$ are both mapped to the element $z$ by the operator $A$, one finds the validity of the evaluation
\begin{equation}
\norm{B\inv{(AB)}z}
	\geq \norm{\inv{A}z}
\end{equation}
from the minimality property of the partial inverse (or more precisely, the validity of the equality
\begin{equation}\label{eq:decomposition_composition_mid}
\norm{B\inv{(AB)}z}^{2}
	= \norm{B\inv{(AB)}z - \inv{A}z}^{2} + \norm{\inv{A}z}^{2}
\end{equation}
from \eqref{eq:decomposition_minimality} regarding the partial inverse of $A$).  The uniqueness of the argument of the minimum (or more directly, the equality \eqref{eq:decomposition_composition_mid}) thus reveals the equivalence of the conditions
\begin{equation}\label{char:partial-inverse_clean_2}
B\inv{(AB)}z = \inv{A}z
	\iff \norm{B\inv{(AB)}z} = \norm{\inv{A}z}.
\end{equation}

In view of the fact that the condition $z \in \dom{\inv{(AB)}}$ and $\inv{(AB)}z = \inv{B}\inv{A}z$ is equivalent to that of $z \in \dom{\inv{B}\inv{A}}$ and $B\inv{(AB)}z = \inv{A}z$, one finally finds by combining \eqref{char:partial-inverse_clean_1} and \eqref{char:partial-inverse_clean_2} the equivalence of the conditions
\begin{enumerate}[label=\rm{(\alph*)}]
\item $\inv{(AB)}z = \inv{B}\inv{A}z$,\label{char:partial-inverse_composition_local_1}
\item $\norm{\inv{(AB)}z} = \norm{\inv{B} \inv{A} z}$,\label{char:partial-inverse_composition_local_2}
\item $B\inv{(AB)}z = \inv{A}z$,\label{char:partial-inverse_composition_local_3}
\item $\norm{B\inv{(AB)}z} = \norm{\inv{A}z}$,\label{char:partial-inverse_composition_local_4}
\end{enumerate}
regarding the partial inverse of composite operators.
A simple example in which the above equivalent conditions are satisfied is when the operator $B$ is an isometry.  Indeed, given that the partial inverse of an isometry is itself an isometry by construction, one finds the evaluation $\norm{ \inv{(AB)}z } \leq \norm{ \inv{B} \inv{A} z } = \norm{ \inv{A} z } \leq \norm{ B\inv{(AB)}z } = \norm{ \inv{(AB)}z }$, thereby fulfilling the aforementioned conditions as desired.

\subsection{Condition regarding Boundedness\label{sec:partial-inverse:boundedness}}

As bounded operators are closed, their partial inverses are always well-defined.  Meanwhile, boundedness of an operator are not necessarily transferred to its partial inverse; the partial inverse of a bounded operator may be unbounded.

The characterisation of the condition on which the partial inverse $\inv{A}$ of a closed operator $A$ is bounded is dictated by the renowned closed graph theorem;  given that $\inv{A}$ is closed as an operator, one finds form a straightforward application of the theorem that $\inv{A}$ is bounded if and only if its domain is topologically closed, or equivalently, if and only if the closed operator $A$ has a closed range, \textit{i.e.}, $\cran{A} = \ran{A}$.

\subsection{Adjoint}

If the closed operator $A$ happens to be densely defined, its adjoint becomes meaningful.  In such a case, one immediately finds
\begin{equation}
D_{\!A} = \cran{A^{*}} \cap \dom{A},
\end{equation}
given the basic fact from functional analysis that the orthogonal complement $\ker{A}^{\perp} = \cran{A^{*}}$ of the kernel of a densely defined closed operator $A$ is equal to the topological closure of the range of the adjoint.  It is then fairly straightforward to see that $D_{\!A}$ is dense in the closed subspace $\cran{A^{*}}$.

It now becomes tempting to consider the adjoint of the partial inverse.  One simple way to attain this is to restrict the Hilbert space encompassing the domain $\dom{\inv{A}} = \ran{A}$ of the partial inverse to its completion so that the map \eqref{def:partial-inverse_restriction} may be understood as a densely defined Hilbert-space operator in its own right:
\begin{equation}\label{def:partial-inverse_restriction}
\inv{A} : \cran{A} \supset \dom{\inv{A}} \to \HS{H}.
\end{equation}
One may then verify using the rudimentary techniques of functional analysis that the partial inverse
\begin{equation}\label{eq:partial-inverse_adjoint}
\inv{(A^{*})} = (\inv{A})^{*}
\end{equation}
of the adjoint of a closed operator coincides with the adjoint of its partial inverse.

In consideration of the fact that the products \eqref{char:partial-inverse_01} and \eqref{char:partial-inverse_02} of the operators concerned are bounded densely defined symmetric operators, their adjoints are nothing but their unique continuous extensions;  explicitly, they respectively read $(\inv{A}A)^{*} = P_{\cran{A^{*}}}$ and $(A\inv{A})^{*} = \id_{\cran{A}}$, where $P_{\cran{A^{*}}}$ is the orthogonal projection associated with the closed subspace $\cran{A^{*}}$, and $\id_{\cran{A}}$ is the identity operator on $\cran{A}$.  These facts reveal
\begin{align}
(\inv{A}A)^{**} = (\inv{A}A)^{*} &\supset \inv{A}A, \\
(A\inv{A})^{**} = (A\inv{A})^{*} &\supset A\inv{A},
\end{align}
which may be understood as a generalisation of the remaining two characterising conditions of the familiar `general reciprocal' \cite{Moore_1920, Bjerhammar_1951_01, Bjerhammar_1951_02, Bjerhammar_1958, Penrose_1955} defined for finite-dimensional matrices.

\section{Losses of a Process\label{sec:loss}}

The pullback and the pushforward, along with their partial inverses, furnish the basis of the definitions of the loss of a process.  As before, in order to avoid unnecessary repetition, two of the most relevant types of processes, namely, quantum measurements and quantum processes, are given special focus below.  It is to be stressed again that, as mentioned earlier, parallel definitions, properties, and facts described hereafter are also valid for any other types of processes;  their mathematical description, operational interpretations, and physical ramifications shall be addressed elsewhere by the author in an appropriate context.

\subsection{Loss of a Process}

Under the tools and concepts introduced so far, the author defines the \emph{loss} of a process regarding an observable over a state by the amount of contraction induced by the pushforward.  As mentioned above, the current paper focuses primarily on the description of the loss of quantum measurements and quantum processes, which shall be respectively termed as \emph{error} and \emph{disturbance} in this paper.  Obvious parallel definitions are available for any other processes, which shall be given appropriate names and thorough descriptions elsewhere in the subsequent papers of the author in suitable contexts.

\subsubsection{Error and Disturbance}

Under the tools and concepts introduced so far, the author defines the \emph{error} of an observable $A$ regarding a quantum measurement $M$ over a state $\rho$ by the amount of contraction
\begin{equation}\label{def:error_quantum-measurement}
\err{A}{M}{\rho}
	\defeq \sqrt{\lnorm{A}{\rho}^{2} - \lnorm{\pf{M}{\rho}A}{M\!\rho}^{2}}
\end{equation}
induced by the pushforward.  In a parallel manner, the author defines the \emph{disturbance} of an observable $A$ regarding a quantum process $\Theta$ over a state $\rho$ by the amount of contraction
\begin{equation}\label{def:disturbance_quantum-process}
\dst{A}{M}{\rho}
	\defeq \sqrt{\lnorm{A}{\rho}^{2} - \lnorm{\pf{\Theta}{\rho}A}{\Theta\hspace{-0.05em}\rho}^{2}}
\end{equation}
induced by the pushforward.

Here, the error and disturbance are defined in such a way that both are of essentially the same nature, each representing the amount of `loss' induced by the respective processes, namely, the quantum measurement $M$ (\textit{i.e.}, quantum-to-classical process) and the quantum process $\Theta$ (\textit{i.e.}, quantum-to-quantum process).  Again, it goes without saying that, in general, this method of quantifying the loss by the amount of contraction is applicable to any types of processes, including classical measurements (classical processes) and classical-to-quantum processes, in obvious manners.

The error \eqref{def:error_quantum-measurement} and disturbance \eqref{def:disturbance_quantum-process} each furnish seminorms on the space $\qob{H}{\rho}$ of localised quantum observables, which could be depicted as a `tangent' space attached to the point $\rho$ over which the processes are applied;  more explicitly, non-negativity $\err{A}{M}{\rho} \geq 0$ of the error follows immediately from the non-expansiveness of the pushforward, whereas the absolute homogeneity $\err{tA}{M}{\rho} = \abs{ t }\, \err{A}{M}{\rho}$, $\forall t \in \mathbb{R}$, and the subadditivity $\err{A}{M}{\rho} + \err{B}{M}{\rho} \geq \err{A+B}{M}{\rho}$ may also be confirmed straightforwardly (the same properties hold for the disturbance $\dst{A}{\Theta}{\rho}$ as well).

\subsubsection{Characterisation of the Error}

The error \eqref{def:error_quantum-measurement} admits an operational interpretation as the minimal cost of the local reconstruction of the quantum observable $A$ through the measurement $M$ over the state $\rho$.  To expound on this, let the act of local reconstruction be implemented by the pullback $\dpb{M}{\rho}$ of the measurement, which creates localised quantum observables $\dpb{M}{\rho} f \in \qob{H}{\rho}$ out of localised classical observables $f \in \cob{\Omega}{M\!\rho}$.  The precision of the local reconstruction shall be evaluated in this paper by the gauge
\begin{equation}\label{def:gauge}
\varepsilon_{\rho}(A,f;M)
	\defeq \sqrt{ \lnorm{ A - \dpb{M}{\rho}f }{\rho}^{2} + \Bigl( \lnorm{ f }{M\!\rho}^{2} - \lnorm{ \dpb{M}{\rho}f }{\rho}^{2} \Bigr) },
\end{equation}
which may itself be interpreted as another definition of error involving classical observables.  Here, the first term $\lnorm{ A - \dpb{M}{\rho}f }{\rho}$ gives an evaluation of the algebraic deviation between the target observable $A$ to be reconstructed and the operator $\dpb{M}{\rho}f$ created by means of the measurement $M$ out of $f$, whereas the second term $\lnorm{ f }{M\!\rho}^{2} - \lnorm{ \dpb{M}{\rho}f }{\rho}^{2} = \lstdv{f}{M\!\rho}^{2} - \lstdv{\dpb{M}{\rho}f}{M\!\rho}^{2} \geq 0$ represents the potential increase of the cost in the reconstruction itself owing to the suboptimal choice of either or both of the measurement $M$ and the function $f$ (see discussions below the equality \eqref{ineq:non-expansiveness_adjoint}).

A useful observation is that the square of the gauge \eqref{def:gauge} admits the decomposition
\begin{equation}\label{eq:decomposition_gauge}
\varepsilon_{\rho}(A,f;M)^{2}
	= \err{A}{M}{\rho}^{2} + \lnorm{ \dpf{M}{\rho}A - f }{M\!\rho}^{2}
\end{equation}
into the sum of the squares of the error \eqref{def:error_quantum-measurement} and the potential cost induced by the suboptimality of the choice of $f$;  this could be confirmed by simple computation using \eqref{char:pullback_pushforward}.

At this point, one immediately finds that the classical observable that minimises the equality \eqref{eq:decomposition_gauge} is furnished by the pushforward $f = \dpf{M}{\rho}A$;  this offers an operational characterisation of the error
\begin{equation}\label{char:error}
\min_{f} \varepsilon_{\rho}(A,f;M)
	= \err{A}{M}{\rho}
\end{equation}
as the minimum of the gauge over all $f$ in reconstructing the quantum observable $A$, along with the interpretation of the pushforward as the unique (up to equivalence) optimal function that attains it.

It is also worth noting that, in view of the decomposition $\err{A}{M}{\rho}^{2} = \lstdv{A}{\rho}^{2} - \lstdv{\pf{M}{\rho}A}{M\!\rho}^{2}$, the error is always bounded from above by the standard deviation
\begin{equation}
\lstdv{A}{\rho} \geq \err{A}{M}{\rho} \geq 0,
\end{equation}
and is therefore bounded with respect to the seminorm $\lnorm{A}{\rho}$ on $\qob{H}{\rho}$, hence is continuous.

\subsubsection{Characterisation of the Disturbance}

The disturbance \eqref{def:disturbance_quantum-process} also admits an operational interpretation in terms of the error \eqref{def:error_quantum-measurement} regarding composite quantum measurements.  To expound on this, let $\Theta : Z(\mathcal{H}) \to Z(\mathcal{K})$ be a quantum process, and let $L : Z(\mathcal{K}) \to W(\Omega)$ be any quantum measurement performed on the resultant quantum-state space $Z(\mathcal{K})$.  The square of the error of the composite quantum measurement $L \circ \Theta$ is readily found to admit a decomposition
\begin{align}\label{eq:decomposition_error_composite-measurement}
&\err{A}{L \circ \Theta}{\rho}^{2} \notag \\
	&\qquad= \lnorm{ A }{\rho}^{2} - \lnorm{ \spf{(L \circ \Theta)} A }{(L \circ \Theta)\rho}^{2} \notag \\
	&\qquad= \lnorm{ A }{\rho}^{2} - \lnorm{ \spf{\Theta} A }{\Theta\hspace{-0.05em}\rho}^{2} + \lnorm{ \spf{\Theta} A }{\Theta\hspace{-0.05em}\rho}^{2} - \lnorm{ \spf{L} (\spf{\Theta} A) }{L (\Theta\hspace{-0.05em}\rho)}^{2} \notag \\
	&\qquad= \dst{A}{\Theta}{\rho}^{2} + \err{\spf{\Theta}A}{L}{(\Theta\hspace{-0.05em}\rho)}^{2}
\end{align}
into the sum of the square of the disturbance of the initial process $\Theta$ and that of the error of the secondary measurement $L$, where the composition law \eqref{eq:composition_pushforward} of the pushforward is employed in the second equality.  Here, and also in what follows, the abbreviated notations for the pullback $\spb{\Theta} = \pb{\Theta}{\rho}$ and the pushforward $\spf{\Theta} = \pf{\Theta}{\rho}$ (and similarly $\spf{M} = \pf{M}{\rho}$ and $\spb{M} = \pb{M}{\rho}$ for quantum measurements, as well as those for any other processes) are occasionally adopted to avoid unnecessary clutter whenever the states over which the process is performed is obvious by context.

A trivial corollary to the decomposition \eqref{eq:decomposition_error_composite-measurement} is the evaluation
\begin{equation}
\err{A}{L \circ \Theta}{\rho}
	\geq \dst{A}{\Theta}{\rho}
\end{equation}
valid for any quantum measurement $L$ performed over the outcome space $\mathcal{K}$ of the quantum process $\Theta$.  Given that every self-adjoint observable admits a (global) errorless measurement by its projection measurement, one has, by construction of the space $S_{\rho}(\mathcal{H})$, an operational characterisation of the disturbance
\begin{equation}\label{char:disturbance}
\inf_{L} \err{A}{L \circ \Theta}{\rho}
	= \dst{A}{\Theta}{\rho}
\end{equation}
as the infimum of the errors of composite quantum measurements, along with the interpretation of the pushforward $\spf{\Theta}A$ as the indicator of the (sequence of) locally optimal choices of the secondary measurement(s) that attains the infimum (in the limit).

\subsection{Loss of a Process under Local Representability}

In this paper, an observable $A \in \qob{H}{\rho}$ is said to be \emph{locally representable} by a quantum measurement $M : Z(\HS{H}) \to W(\Omega)$ over the state $\rho$ (or briefly, $\rho$-representable), if the observable $A \in \ran{\pb{M}{\rho}}$ belongs to the image of the pullback;  a classical observable $f \in \cob{\Omega}{M\!\rho}$ satisfying $\pb{M}{\rho}f = A$ shall then be called a \emph{local representative} of $A$ regarding $M$ over $\rho$ (or briefly, a $\rho$-representative).
In a parallel manner, an observable $A \in \qob{H}{\rho}$ is said to be \emph{locally representable} by a quantum process $\Theta : Z(\HS{H}) \to Z(\HS{K})$ over the state $\rho$ (or briefly, $\rho$-representable), if the observable $A \in \ran{\pb{\Theta}{\rho}}$ belongs to the image of the pullback;  a quantum observable $C \in \qob{K}{(\Theta\hspace{-0.05em}\rho)}$ satisfying $\pb{M}{\rho}C = A$ shall then be called a \emph{local representative} of $A$ regarding $M$ over $\rho$ (or briefly, a $\rho$-representative).

\subsubsection{Error and Disturbance under Local Representability}

The concept of local representability leads to another natural definition of the error and disturbance.  The author defines the \emph{error for local representability}
\begin{equation}\label{def:error_quantum-measurement_representability}
\rerr{A}{M}{\rho}
	\defeq \sqrt{\lnorm{\invpb{M}{\rho}A}{M\!\rho}^{2} - \lnorm{A}{\rho}^{2}}
\end{equation}
of a measurement $M$ of a locally representable observable $A \in \ran{\pb{M}{\rho}}$ over $\rho$ by the amount of expansion induced by the partial inverse \eqref{def:partial-inverse} of the pullback of $M$.  In a parallel manner, the author defines the \emph{disturbance for local representability}
\begin{equation}\label{def:disturbance_quantum-process_representability}
\rdst{A}{M}{\rho}
	\defeq \sqrt{\lnorm{\invpb{\Theta}{\rho}A}{\Theta\hspace{-0.05em}\rho}^{2} - \lnorm{A}{\rho}^{2}}
\end{equation}
of a quantum process $\Theta$ of a locally representable observable $A \in \ran{\pb{\Theta}{\rho}}$ over $\rho$ by the amount of expansion induced by the partial inverse \eqref{def:partial-inverse} of the pullback of $\Theta$.  Given the boundedness---and hence the closedness---of the pullbacks, their partial inverses are well-defined closed operators, and consequently the error \eqref{def:error_quantum-measurement_representability} and disturbance \eqref{def:disturbance_quantum-process_representability} are both well-defined quantities.

Note that, while the pullback is always bounded, its partial inverse is in general not necessarily so, and as a consequence, the error \eqref{def:error_quantum-measurement_representability} and disturbance \eqref{def:disturbance_quantum-process_representability} may be unbounded;  from the characterising condition on which the partial inverse of a closed operator is bounded (see Sec.~\ref{sec:partial-inverse:boundedness}), such are the cases if and only if the domains of partial inverses of the pullbacks are topologically closed, or equivalently, if and only if the pullbacks have closed range (\textit{i.e.}, $\ran{\pb{M}{\rho}} = \cran{\pb{M}{\rho}}$ and $\ran{\pb{\Theta}{\rho}} = \cran{\pb{\Theta}{\rho}}$).

As with the error \eqref{def:error_quantum-measurement} and disturbance \eqref{def:disturbance_quantum-process} introduced earlier, the error \eqref{def:error_quantum-measurement_representability} and disturbance \eqref{def:disturbance_quantum-process_representability} under local representability each furnish seminorms on the space $\qob{H}{\rho}$ of localised quantum observables;  non-negativity $\rerr{A}{M}{\rho} \geq 0$ of the error \eqref{def:error_quantum-measurement_representability} for local representability follows from the non-expansiveness of the pushforward combined with the property \eqref{char:partial-inverse_02} of the partial inverse, whereas the absolute homogeneity $\rerr{tA}{M}{\rho} = \abs{t}\, \rerr{A}{M}{\rho}$, $\forall t \in \mathbb{R}$, and the subadditivity $\rerr{A}{M}{\rho} + \rerr{B}{M}{\rho} \geq \rerr{A+B}{M}{\rho}$ may also be confirmed straightforwardly (the same properties hold for the disturbance $\dst{A}{\Theta}{\rho}$ as well).

\subsubsection{Closedness of the Loss}

Closedness of the partial inverses as Hilbert-space operators entails closedness of the error \eqref{def:error_quantum-measurement_representability} and disturbance \eqref{def:disturbance_quantum-process_representability} as (the square roots of) quadratic forms on Hilbert spaces, which are noteworthy and convenient property of the novel construction of error and disturbance presented in this paper.

Recall that a symmetric (\textit{alias} Hermitian) sesquilinear form (or briefly, a symmetric form) $\mathfrak{t}$ defined on a subspace $D_{\mathfrak{t}}$ of a Hilbert space $\HS{H}$ is called lower semibounded, if there exists a real number $m \in \mathbb{R}$ such that $\mathfrak{t}[x,x] \geq m \norm{x}^{2}$ holds for all $x \in D_{\mathfrak{t}}$, where $\norm{\,\cdot\,}$ is the norm of the original Hilbert space $\HS{H}$.  In such a case, the number $m$ is called a lower bound for $\mathfrak{t}$, and it is customary to write $\mathfrak{t} \geq m$.  If the form has $0$ as a lower bound, \textit{i.e.}, $\mathfrak{t} \geq 0$, it is called positive.

A lower semibounded symmetric form $\mathfrak{t}$ defined on a dense subspace $D_{\mathfrak{t}}$ of a Hilbert space is called closed, if the space $D_{\mathfrak{t}}$ is complete with respect to the inner product $\linpr{x}{y}{\mathfrak{t}} \defeq \mathfrak{t}[x,y] + (1 - m) \inpr{x}{y}$, where $\inpr{\,\cdot\,}{\,\cdot\,}$ is the inner product of the original Hilbert space $\HS{H}$.

Given the fact that the error \eqref{def:error_quantum-measurement_representability} and disturbance \eqref{def:disturbance_quantum-process_representability} for local representability fulfil the parallelogram identity, which could be readily confirmed, they induce positive symmetric forms uniquely associated with them.  With rudimentary techniques of functional analysis, it is then straightforward to find by construction that the closedness of the partial inverses entails the closedness of the positive symmetric forms uniquely associated with the error \eqref{def:error_quantum-measurement_representability} and disturbance \eqref{def:disturbance_quantum-process_representability}.

\subsubsection{Lower Semicontinuity of the Loss}

As is basic, a useful characterisation of closedness of forms may be given in terms of lower semicontinuity.  Recall that a function $f : X \to \mathbb{R} \cup \{\infty\}$ on a metric space $X$ is called lower semicontinuous, if $\liminf_{n\to\infty} f(x_{n}) \geq f(\lim_{n\to\infty} x_{n})$ holds for any convergent sequence $x_{n}$ in $X$.  Under the same settings as above, with a slight abuse of notation, let $\mathfrak{t}[x] \defeq \mathfrak{t}[x,x]$ denote the quadratic form associated with a densely defined lower semibounded symmetric form $\mathfrak{t}$, and consider its extension $\mathfrak{t}^{\prime} : \HS{H} \to \mathbb{R} \cup \{\infty\}$ to the whole Hilbert space by letting $\mathfrak{t}^{\prime}[x] \defeq \mathfrak{t}[x]$ for $x \in D_{\mathfrak{t}}$ and $\mathfrak{t}^{\prime}[x] \defeq +\infty$ otherwise.  Then, it is well-known that $\mathfrak{t}$ is closed if and only if its extension $\mathfrak{t}^{\prime}$ is lower semicontinuous.

Under the current context of quantum measurements and quantum processes, one may thus extend the original definitions of error \eqref{def:error_quantum-measurement_representability} and disturbance \eqref{def:disturbance_quantum-process_representability} for local representability to the completion (or equivalently, the topological closure) of the original domain by defining
\begin{equation}\label{def:extension_error_quantum-measurement_representability}
\rerr{A}{M}{\rho}
	\defeq
		\begin{dcases}
		\rerr{A}{M}{\rho} & (A \in \ran{\pb{M}{\rho}}), \\
		+\infty & (A \notin \ran{\pb{M}{\rho}}),
		\end{dcases}
\end{equation}
and
\begin{equation}\label{def:extension_disturbance_quantum-process_representability}
\rdst{A}{\Theta}{\rho}
	\defeq
		\begin{dcases}
		\rdst{A}{\Theta}{\rho} & (A \in \ran{\pb{\Theta}{\rho}}), \\
		+\infty & (A \notin \ran{\pb{\Theta}{\rho}}),
		\end{dcases}
\end{equation}
where a slight abuse of notations is made in which the original definitions and their extensions are denoted by the same symbols.

Given the closedness of the error \eqref{def:error_quantum-measurement_representability} and disturbance \eqref{def:disturbance_quantum-process_representability} as (the square roots of) quadratic forms, it is then immediate from the aforementioned characterisation that the extended definitions \eqref{def:extension_error_quantum-measurement_representability} and \eqref{def:extension_disturbance_quantum-process_representability} are lower semicontinuous;  more explicitly, for any convergent sequence $A_{n} \to A$ of localised quantum observables in the extended domain, the inequalities
\begin{equation}
\liminf_{n\to\infty} \rerr{A_{n}}{M}{\rho}
	\geq \rerr{A}{M}{\rho}
\end{equation}
and
\begin{equation}
\liminf_{n\to\infty} \rdst{A_{n}}{\Theta}{\rho}
	\geq \rdst{A}{\Theta}{\rho}
\end{equation}
hold.  The extended definitions \eqref{def:extension_error_quantum-measurement_representability} and \eqref{def:extension_disturbance_quantum-process_representability} thus respectively furnish the lower bounds to any convergent subsequences of the errors $\rerr{A_{n_{k}}}{M}{\rho}$ and disturbances $\rdst{A_{n_{k}}}{\Theta}{\rho}$ associated with the subsequence $A_{n_{k}}$ of the original convergent sequence $A_{n}$ of localised quantum observables.

While the following passages of this paper focus mostly on the original definitions of the error \eqref{def:error_quantum-measurement_representability} and disturbance \eqref{def:disturbance_quantum-process_representability} for the sake of simplicity of arguments, many of the results and properties presented there remain valid (with due reformulations in obvious manners if necessary) for the extended definitions \eqref{def:extension_error_quantum-measurement_representability} and \eqref{def:extension_disturbance_quantum-process_representability} as well with the usual mathematical conventions regarding the infinity.  Further extension of the definitions \eqref{def:error_quantum-measurement_representability} and \eqref{def:disturbance_quantum-process_representability} to the whole Hilbert space $\qob{H}{\rho}$ is also possible;  it will be addressed elsewhere in an appropriate context.

\subsubsection{Characterisation of the Error under Local Representability}

As with the error \eqref{def:error_quantum-measurement}, the error \eqref{def:error_quantum-measurement_representability} for local representability also admits an operational interpretation as the minimal cost of the local reconstruction of the quantum observable $A \in \ran{\pb{M}{\rho}}$, albeit under a different constraint.

A useful observation for the argument is that, under the local-representability condition $\pb{M}{\rho}f = A$ of the measurement, the square of the gauge \eqref{def:gauge} admits a decomposition
\begin{equation}\label{eq:decomposition_gauge_representability}
\varepsilon_{\rho}(A,f;M)^{2} = \rerr{A}{M}{\rho}^{2} + \lnorm{ \invpb{M}{\rho}A - f }{M\!\rho}^{2}
\end{equation}
into the sum of the squares of the error \eqref{def:error_quantum-measurement_representability} and the potential cost of the suboptimal choice of the local representative;  this may be readily confirmed by simple computation utilising the construction and the minimality property \eqref{char:minimality} of the partial inverse.

At this point, it is easy to see that the $\rho$-representative that minimises the gauge \eqref{eq:decomposition_gauge_representability} is furnished by the partial inverse $f = \invpb{M}{\rho} A$ regarding the quantum observable of interest;  this offers an operational characterisation of the error
\begin{equation}\label{char:error_representability}
\min_{\pb{M}{\rho}f = A} \varepsilon_{\rho}(A,f;M)
	= \rerr{A}{M}{\rho}
\end{equation}
as the minimum of the gauge over all the classical observables $f$ in representing the quantum observable $A$ under the constraint $\pb{M}{\rho}f = A$ of local representability, along with the interpretation of the partial inverse as the unique (up to equivalence) locally optimal function that attains it.

For another explicit interpretation (compare the results below with those of estimation theory \cite{Rao_1945,Cramere_1946,Holevo_1982}, regarding which the details shall be given more thoroughly in subsequent publications of the author), given the fact that the expectation value is invariant under the pullback and the pushforward as mentioned earlier (it shall be also noted here that, as a corollary to this property, the partial inverses of the pullback $\lev{ \invpb{M}{\rho} A }{M\!\rho} = \lev{A}{\rho}$ and the pushforward $\lev{ \invpf{M}{\rho} f }{\rho} = \lev{f}{M\!\rho}$ also preserve the expectation value), observe that the equality \eqref{eq:decomposition_gauge_representability} may be rewritten into
\begin{equation}\label{eq:decomposition_standard-deviation_representability}
\lstdv{f}{M\!\rho}^{2}
	=  \lstdv{A}{\rho}^{2} + \rerr{A}{M}{\rho}^{2} + \lnorm{ \invpb{M}{\rho}A - f }{M\!\rho}^{2}.
\end{equation}
This admits an interpretation as a decomposition of the variance of a $\rho$-representative $f$ of an observable $A$ into the sums of the squares of the quantum fluctuation (quantum standard deviation) of $A$, that of the error \eqref{def:error_quantum-measurement_representability}, and that of the contribution of the suboptimal choice of the function $f$.

Specifically, note that the first component $\lstdv{A}{\rho}$ of the decomposition \eqref{eq:decomposition_standard-deviation_representability} is determined solely by the choice of the quantum observable and the state, thereby setting a universal lower bound to the statistical cost $\lstdv{f}{M\!\rho}$ of the reconstruction of the quantum observable $A$ of interest, whereas the second component $\rerr{A}{M}{\rho}$ contributes to the additional fundamental bound imposed by the choice of the measurement $M$.  In this regard, it is beneficial to note a trivial corollary
\begin{equation}\label{ev:standard-deviation_lower_bound}
\lstdv{f}{M\!\rho}^{2} \geq \lstdv{A}{\rho}^{2} + \rerr{A}{M}{\rho}^{2}
\end{equation}
to the decomposition \eqref{eq:decomposition_standard-deviation_representability} in the form of an inequality, in which the role of the two contributors to the lower bound becomes more transparent.

The third component $\lnorm{ \invpb{M}{\rho}A - f }{M\!\rho}$ of the decomposition \eqref{eq:decomposition_standard-deviation_representability} furthermore takes into account the additional cost induced by the suboptimal choice of the local representative $f$.  As is familiar, the choice of the function $f$ effectively corresponds to the choice of the method of processing the measurement outcomes, and as such is often the primary place in which one seeks for optimisation in many real situations.  The decomposition \eqref{eq:decomposition_standard-deviation_representability} dictates that the optimal choice of the representative is uniquely given by the partial inverse $f = \invpb{M}{\rho} A$, in which case the decomposition  reads 
\begin{equation}
\dlstdv{\invpb{M}{\rho} A}{M\!\rho}^{2} = \lstdv{A}{\rho}^{2} + \rerr{A}{M}{\rho}^{2}
\end{equation}
and the lower bound \eqref{ev:standard-deviation_lower_bound} of the statistical cost is uniquely attained.

\subsubsection{Characterisation of the Disturbance under Local Representability}

As with the disturbance \eqref{def:disturbance_quantum-process}, the disturbance \eqref{def:disturbance_quantum-process_representability} for local representability also admits an operational interpretation in terms of the error \eqref{def:error_quantum-measurement_representability} regarding composite quantum measurements.

To expound on this, let $\Theta : Z(\mathcal{H}) \to Z(\mathcal{K})$ be a quantum process, and let $A \in \ran{\pb{\Theta}{\rho}}$ be a $\rho$-representable quantum observable.  On top of this, let $L : Z(\mathcal{K}) \to W(\Omega)$ be a quantum measurement performed on the resultant quantum-state space $Z(\mathcal{K})$ such that $A \in \ran{\pb{(L \circ \Theta)}{\rho}}$ is $\rho$-representable with respect to the composite measurement $L \circ \Theta$.
One then finds, using similar techniques as the derivation of \eqref{eq:decomposition_error_composite-measurement} along with \eqref{eq:decomposition_composition}, the decomposition
\begin{align}\label{eq:decomposition_error_composite-measurement_representability_1}
\rerr{A}{L\circ\Theta}{\rho}^{2}
	&= \rdst{A}{\Theta}{\rho}^{2} + \rerr{\inv{ (\spb{\Theta}) } A}{L}{(\Theta\hspace{-0.05em}\rho)}^{2}  \notag \\
	&\quad - \lnorm{ \inv{ (\spb{L}) } \inv{ (\spb{\Theta}) } A - \inv{ (\spb{\Theta} \spb{L}) } A }{\rho}^{2}
\end{align}
of the square of the error of the composite measurement into the contributions of the square of the disturbance of the initial process $\Theta$ on the quantum observable $A$, that of the error of the secondary measurement $L$ of the partial inverse $\inv{ (\spb{\Theta}) } A$, and that of the deviation $\lnorm{ \inv{ (\spb{L}) } \inv{ (\spb{\Theta}) } A - \inv{ (\spb{\Theta} \spb{L}) } A }{\rho}^{2} = \lnorm{ \inv{ (\spb{L}) } \inv{ (\spb{\Theta}) } A }{\rho}^{2} - \lnorm{ \inv{ (\spb{\Theta} \spb{L}) } A }{\rho}^{2}$ of the element $\inv{ (\spb{L}) } \inv{ (\spb{\Theta}) } A$ from the partial inverse $\inv{ (\spb{\Theta} \spb{L}) } A = \inv{ (\spb{N}) } A$ regarding the composite quantum measurement $N \defeq L \circ \Theta$.

Using parallel techniques as above, this time with \eqref{eq:decomposition_composition_mid} instead of \eqref{eq:decomposition_composition}, one finds another decomposition
\begin{align}\label{eq:decomposition_error_composite-measurement_representability_2}
\rerr{A}{L\circ\Theta}{\rho}^{2}
	= \rdst{A}{\Theta}{\rho}^{2} &+ \rerr{\Theta_{L} A}{L}{(\Theta\hspace{-0.05em}\rho)}^{2}  \notag \\
	&+ \lnorm{ \Theta_{L} A - \inv{ (\spb{\Theta}) } A }{\rho}^{2}
\end{align}
of the square of the error into the sum of the square of the disturbance of the initial process $\Theta$ of $A$, that of the error of the secondary measurement $L$ of the element $\Theta_{L} A$ with the shorthand $\Theta_{L} \defeq \spb{L} \inv{ (\spb{\Theta} \circ \spb{L}) }$, and that of the deviation $\lnorm{ \Theta_{L} A - \inv{ (\pb{\Theta}{\rho}) } A }{\rho}^{2} = \lnorm{ \Theta_{L} A }{\rho}^{2} - \lnorm{ \inv{ (\pb{\Theta}{\rho}) } A}{\rho}^{2}$ of the element $\Theta_{L} A$ from the partial inverse $\inv{ (\pb{\Theta}{\rho}) } A$ regarding the quantum process $\Theta$.

The two results \eqref{eq:decomposition_error_composite-measurement_representability_1} and \eqref{eq:decomposition_error_composite-measurement_representability_2} entail as a trivial corollary the evaluation of the error of the composite quantum measurement
\begin{align}\label{eq:evaluation_error_composite-measurement_representability}
&\rdst{A}{\Theta}{\rho}^{2} + \rerr{\inv{(\spb{\Theta})} A}{L}{(\Theta\hspace{-0.05em}\rho)}^{2} \notag \\
	&\qquad \geq \rerr{A}{L\circ\Theta}{\rho}^{2} \notag \\
	&\qquad \qquad \geq \rdst{A}{\Theta}{\rho}^{2} + \rerr{\Theta_{L} A}{L}{(\Theta\hspace{-0.05em}\rho)}^{2},
\end{align}
in which both the upper and lower bounds are dictated by the sums of the squares of the disturbance of the quantum process $\Theta$ and the error of the secondary measurement $L$ (note that this trivially entails
\begin{equation}
\rerr{A}{L\circ\Theta}{\rho}
	\geq \rdst{A}{\Theta}{\rho}
\end{equation}
by ignoring the error in the second inequality).  The arguments found in Sec.~\ref{sec:partial-inverse:composition} then reveal the equivalences
\begin{enumerate}[label=\rm{(\alph*)}]
\item The first equality of the evaluation \eqref{eq:evaluation_error_composite-measurement_representability} holds,
\item The second equality of the evaluation \eqref{eq:evaluation_error_composite-measurement_representability} holds,
\item Both the equalities of the evaluation \eqref{eq:evaluation_error_composite-measurement_representability} hold,
\item $\inv{ (\spb{\Theta} \circ \spb{L}) } A = \inv{ (\spb{L}) } \inv{ (\spb{\Theta}) } A$,
\item $\lnorm{ \inv{ (\spb{\Theta} \circ \spb{L}) } A }{(L\circ\Theta)\rho} = \lnorm{ \inv{ (\spb{L}) } \inv{ (\spb{\Theta}) } A }{(L\circ\Theta)\rho}$,
\item $\inv{(\spb{\Theta})} A = \spb{L} \inv{ (\spb{\Theta} \circ \spb{L}) }A $,
\item $\lnorm{ \inv{(\spb{\Theta})} A }{\Theta\hspace{-0.05em}\rho} = \lnorm{ \spb{L} \inv{ (\spb{\Theta} \circ \spb{L}) }A }{\Theta\hspace{-0.05em}\rho}$,
\end{enumerate}
in which case the error of the composite quantum measurement reduces to
\begin{equation}\label{eq:decomposition_error_composite-measurement_representability}
\rerr{A}{L\circ\Theta}{\rho}^{2}
	= \rdst{A}{\Theta}{\rho}^{2} + \rerr{\inv{(\spb{\Theta})} A}{L}{(\Theta\hspace{-0.05em}\rho)}^{2},
\end{equation}
which may be understood as a parallel relation to \eqref{eq:decomposition_error_composite-measurement}.

An archetypal example in which the relation \eqref{eq:decomposition_error_composite-measurement_representability} holds is when the secondary measurement $L$ is a projection measurement, for indeed the pullback of a projection measurement is an isometry.  As a straightforward corollary to this, given that every self-adjoint observable admits a (global) errorless measurement by its projection measurement, one has an operational characterisation of the disturbance
\begin{equation}\label{char:disturbance_representability}
\inf_{L} \rerr{A}{L\circ\Theta}{\rho}
	= \rdst{A}{\Theta}{\rho}
\end{equation}
as the infimum of the error of composite quantum measurement taken over the secondary measurement $L$, along with the interpretation of the partial inverse $\inv{(\spb{\Theta})}A$ as the indicator of the (sequence of) locally optimal choices of the secondary measurement(s) that attains the infimum (in the limit).

\subsection{Two Definitions of Loss}

The relation between the two definitions \eqref{def:error_quantum-measurement}, \eqref{def:error_quantum-measurement_representability} of the error and those \eqref{def:disturbance_quantum-process}, \eqref{def:disturbance_quantum-process_representability} of the disturbance may be explicitly addressed by the equalities
\begin{equation}\label{eq:two_errors}
\rerr{A}{M}{\rho}^{2} - \err{A}{M}{\rho}^{2}
	= \lnorm{\invpb{M}{\rho}A - \dpf{M}{\rho}A}{\rho}^{2}
\end{equation}
and
\begin{equation}\label{eq:two_disturbances}
\rdst{A}{M}{\rho}^{2} - \dst{A}{M}{\rho}^{2}
	= \lnorm{\invpb{\Theta}{\rho}A - \dpf{\Theta}{\rho}A}{\rho}^{2},
\end{equation}
respectively.  One direct way to confirm this is through a simple application of the characterisation \eqref{char:pullback_pushforward} regarding the adjoints of the pullback and pushforward, and that \eqref{eq:partial-inverse_adjoint} of their partial inverses, the latter of which explicitly reads
\begin{equation}
\invpf{M}{\rho} = \bigl(\invpb{M}{\rho}\bigr)^{*}
\end{equation}
and
\begin{equation}
\invpf{\Theta}{\rho} = \bigl(\invpb{\Theta}{\rho}\bigr)^{*}
\end{equation}
for the respective maps under the current context.

As trivial corollaries, note that the above relations \eqref{eq:two_errors} and \eqref{eq:two_disturbances} respectively entail the evaluations
\begin{equation}\label{eval:two_errors}
\rerr{A}{M}{\rho}
	\geq \err{A}{M}{\rho}
\end{equation}
and
\begin{equation}\label{eval:two_disturbances}
\rdst{A}{\Theta}{\rho}
	\geq \dst{A}{\Theta}{\rho}
\end{equation}
regarding the two definitions.  Another simple way to obtain them is to compare the characterisations \eqref{char:error}, \eqref{char:error_representability} of the two definitions of error, and those \eqref{char:disturbance}, \eqref{char:disturbance_representability} of the two definitions of disturbance.  Regarding the error, the latter \eqref{char:error_representability} involves the constraint $\pb{M}{\rho}f = A$ regarding local representability, whereas the former \eqref{char:error} is free from such constraint in optimising the gauge \eqref{def:gauge};  a similar line of observation reveals the evaluation for the disturbance as well.

\subsection{Lossless Process\label{sec:loss:lossless}}

The identification of the conditions on which the process becomes free from loss is of great importance.  In this paper, a quantum measurement $M$ is said to be capable of an errorless measurement of $A$ over $\rho$, if the error \eqref{def:error_quantum-measurement} and \eqref{def:error_quantum-measurement_representability} vanish (in what follows, one finds that the condition on which the two definitions of error vanish is equivalent, hence this terminology is justified).  Likewise, a quantum process $\Theta$ is said to not disturb the observable $A$ over $\rho$, if the disturbance \eqref{def:disturbance_quantum-process} and \eqref{def:disturbance_quantum-process_representability} vanish (the same remark regarding the equivalence of the condition holds for the disturbance as well).

\subsubsection{Errorless Measurement of an Observable\label{sec:loss:lossless:measurement}}

Among the several characterisations of the errorless measurement, which shall be expounded in detail elsewhere, the following five equivalent conditions that are most relevant to the context of this paper shall be presented below:
\begin{enumerate}[label=\rm{(\alph*)}]
\item $\rerr{A}{M}{\rho} = 0$,\label{def:errorless-measurement_representability}
\item $A = \invpf{M}{\rho}\invpb{M}{\rho}A$,\label{char:errorless-measurement_1}
\item $\rerr{A}{M}{\rho} = \err{A}{M}{\rho}$,\label{char:errorless-measurement_2}
\item $A = \dpb{M}{\rho}\dpf{M}{\rho}A$,\label{char:errorless-measurement_3}
\item $\err{A}{M}{\rho} = 0$.\label{def:errorless-measurement}
\end{enumerate}
Here, a harmless abuse of expression is made in order to avoid clutter, in which obvious preconditions are implicitly assumed;  it goes without saying that, in precise terms, the condition \ref{def:errorless-measurement_representability} should be understood as `$A \in \ran{\pb{M}{\rho}}$ and $\rerr{A}{M}{\rho} = 0$', condition \ref{char:errorless-measurement_1} as `$A \in \ran{\pb{M}{\rho}}$, $\invpb{M}{\rho}A \in \ran{\pf{M}{\rho}}$, and $A = \invpf{M}{\rho}\invpb{M}{\rho}A$', and condition \ref{char:errorless-measurement_2} as `$A \in \ran{\pb{M}{\rho}}$ and $\rerr{A}{M}{\rho} = \err{A}{M}{\rho}$' in order to guarantee their well-definedness.

A simple proof of the above equivalences reads as follows: $\ref{def:errorless-measurement_representability} \implies \ref{def:errorless-measurement}$ is trivial from the evaluation \eqref{eval:two_errors}, $\ref{def:errorless-measurement} \implies \ref{char:errorless-measurement_3}$ is an immediate consequence of the characterisation \eqref{eq:decomposition_gauge}, from which $\err{A}{M}{\rho} \geq \lnorm{ A - \dpb{M}{\rho}\dpf{M}{\rho} A }{\rho}$ follows, $\ref{char:errorless-measurement_3} \implies \ref{char:errorless-measurement_1}$ may be readily confirmed by applying $\invpf{M}{\rho}\invpb{M}{\rho}$ on both-hand sides of the assumption, $\ref{char:errorless-measurement_1} \implies \ref{char:errorless-measurement_2}$ is an immediate consequence of \eqref{eq:two_errors} with a simple substitution of the assumption, and $\ref{char:errorless-measurement_2} \implies \ref{def:errorless-measurement_representability}$ is immediate from \eqref{eq:two_errors}, which yields
\begin{equation}\label{char:errorless-measurement_4}
\invpb{M}{\rho}A = \dpf{M}{\rho}A,
\end{equation}
thereby revealing $\lnorm{A}{\rho} \geq \lnorm{\dpf{M}{\rho}A}{M\!\rho} = \lnorm{\invpb{M}{\rho}A}{M\!\rho} \geq \lnorm{A}{\rho}$ due to the non-expansiveness of (the pullback and) the pushforward and the non-contractivity of their partial inverses (in passing, it may also be of use to note that the condition \eqref{char:errorless-measurement_4} is another necessary and sufficient condition for the two definitions of error to vanish, as one may readily observe by following the logical flow of the above proof).

A consequence worthy of mention is that an errorless measurement of a self-adjoint observable is always available; the projection measurement associated with it is capable of such a measurement \emph{globally}, \textit{i.e.}, over every state $\rho$.  A simple proof would be to observe that the pushforward of an observable $A$ by the projection measurement $M$ associated with it reads $\pf{M}{\rho}A = \mathrm{id}$, where $\mathrm{id}$ is the identity function on the spectrum of $A$.  This fact combined with the condition \ref{char:errorless-measurement_3} shall lead to the desired statement.

\subsubsection{Non-disturbing Quantum Process of an Observable\label{sec:loss:lossless:process}}

Likewise, the condition on which a quantum process $\Theta$ does not disturb a quantum observable $A$ is characterised by the equivalent conditions
\begin{enumerate}[label=\rm{(\alph*)}]
\item $\rdst{A}{\Theta}{\rho} = 0$,\label{def:nondisturbing-process_representability}
\item $A = \invpf{\Theta}{\rho}\invpb{\Theta}{\rho}A$,\label{char:nondisturbing-process_1}
\item $\rdst{A}{\Theta}{\rho} = \dst{A}{\Theta}{\rho}$,\label{char:nondisturbing-process_2}
\item $A = \dpb{\Theta}{\rho}\dpf{\Theta}{\rho}A$,\label{char:nondisturbing-process_3}
\item $\dst{A}{\Theta}{\rho} = 0$,\label{def:nondisturbing-process}
\end{enumerate}
and
\begin{equation}\label{char:nondisturbing-process_4}
\invpb{\Theta}{\rho}A = \dpf{\Theta}{\rho}A,
\end{equation}
the proof of which is parallel to that of the quantum measurement described above.

\subsubsection{Equivalence of the Lossless Conditions of the Two Definitions of Loss}

The above equivalences reveal that the conditions on which the two definitions \eqref{def:error_quantum-measurement} and \eqref{def:error_quantum-measurement_representability} of the error vanish is identical (the same remark holds for the two definitions \eqref{def:disturbance_quantum-process} and \eqref{def:disturbance_quantum-process_representability} of the disturbance as well).  In this regard, while the former definitions \eqref{def:error_quantum-measurement} and \eqref{def:disturbance_quantum-process} are always numerically less than the latter definitions \eqref{def:error_quantum-measurement_representability} and \eqref{def:disturbance_quantum-process_representability} for local representability barring the lossless case, the two definitions can be understood as being equivalent for the purpose of detecting the condition on which the process becomes free from loss.

\section{Uncertainty Relations for Errors under Local Joint-Measurability\label{sec:urel_error}}

Prior to the introduction of the main results, the relations for errors regarding a pair of quantum measurements admitting a local joint-measurement \cite{Lee_urel_2022_01} shall be briefly reviewed below, as they serve as bases for the derivation of those for error and disturbance.  

\subsection{Local Joint-Measurability\label{sec:urel_error:local-joint-measurability}}

The concept of local joint-measurability \cite{Lee_urel_2022_01} is reviewed below, which is a natural but non-trivial extension of the standard notion of (global) joint POVM measurement associated with a pair of `commutative' POVM measurements.

\subsubsection{Joint Measurability}

A pair of quantum measurements $M_{i} : Z(\HS{H}) \to W(\Omega_{i})$, $i = 1, 2$, is said to admit a (global) \emph{joint measurement}, if there exists a mediating quantum measurement $J : Z(\HS{H}) \to W(\Omega_{1} \times \Omega_{2})$ from which both the distributions $M\!\rho$ and $N\!\rho$ are retrieved as marginals from the distribution $J\!\rho$ for all quantum states $\rho \in Z(\HS{H})$;  more explicitly, this entails a pair of classical processes $\pi_{i} : W(\Omega_{1} \times \Omega_{2}) \to W(\Omega_{i})$, $i = 1, 2$, each of which projecting the joint probability distributions to their respective marginals
\begin{align}
(\pi_{1} p)(\omega_{1})
	&\defeq \int_{\Omega_{2}} p(\omega_{1},\omega_{2})\, d\omega_{2} \label{def:projection_1}, \\
(\pi_{2} p)(\omega_{2})
	&\defeq \int_{\Omega_{1}} p(\omega_{1},\omega_{2})\, d\omega_{1} \label{def:projection_2},
\end{align}
thereby satisfying $M_{i} = \pi_{i} \circ J$, $i = 1, 2$.  

\subsubsection{Local Joint-Measurability}

It is straightforward to see by the composition laws of the pullback \eqref{eq:composition_pullback} and that of the pushforward \eqref{eq:composition_pushforward} that (global) joint-measurability of a pair of quantum measurements $M_{i}$, $i = 1, 2$, respectively entail the relations of the pullback
\begin{equation}\label{def:joint-measurability_pb}
\spb{(M_{i})} = \spb{J} \circ \spb{(\pi_{i})}
\end{equation}
and those of the pushforward
\begin{equation}\label{def:joint-measurability_pf}
\spf{(M_{i})} = \spf{(\pi_{i})} \circ \spf{J}
\end{equation}
for each of the measurements viewed as composite quantum measurements.

In regards to the fact that the framework of the author is a \emph{local theory}, it is natural to introduce the weaker notion of \emph{local joint-measurability} extending the standard notion of (global) joint measurability described above.  A pair of quantum measurements $M_{i} : Z(\HS{H}) \to W(\Omega_{i})$, $i = 1, 2$, is said to admit a \emph{local joint-measurement} over the state $\rho \in Z(\HS{H})$, if there exists a mediating quantum measurement $J : Z(\HS{H}) \to W(\Omega_{1} \times \Omega_{2})$ for which either of the two equivalent conditions \eqref{def:joint-measurability_pb} or \eqref{def:joint-measurability_pf} holds regarding the projections;  here, note that the said conditions implicitly connote $M_{i}\rho = (\pi_{i} \circ J)\rho$, which is to say that the probability distributions of the measurement outcomes of $M_{i}$ coincide with that of the composite measurements $\pi_{i} \circ J$ over $\rho$. 

Given that the adjoints of the projections \eqref{def:projection_1} and \eqref{def:projection_2} respectively read $(\pi_{1}^{\prime}f)(\omega_{1},\omega_{2}) = f(\omega_{1})$ and $(\pi_{2}^{\prime}g)(\omega_{1},\omega_{2}) = g(\omega_{2})$, one finds that their pullbacks are both isometries, \textit{i.e.}, $\lnorm{ f }{\pi_{1}p} = \lnorm{ \spb{\pi_{1}}f }{p}$ and $\lnorm{ g }{\pi_{2}p} = \lnorm{ \spb{\pi_{2}}g }{p}$.  This allows for the identifications of the spaces $R_{M\!\rho}(\Omega_{i}) \simeq \spb{\pi_{i}}( R_{M\!\rho}(\Omega_{i}) )$, $i = 1, 2$, regarding each of the measurements with their respective images under the pullbacks, which are subspaces $R_{J\!\rho}(\Omega_{1}\times\Omega_{2})$ of the larger space associated with the outcomes of the mediating measurement $J$ over $\rho$.

It is to be emphasised again that the authors notion of local joint-measurability is in general a much weaker and more flexible notion than the traditional concept of global joint-measurability.
 
\subsection{Uncertainty Relations for Errors under Local Joint-Measurability}

Based on the terminologies introduced above, let $M : Z(\HS{H}) \to W(\Omega_{1})$ and $N : Z(\HS{H}) \to W(\Omega_{2})$ be a pair of quantum measurements, and assume that the pair admits a local joint-measurement by $J : Z(\HS{H}) \to W(\Omega_{1} \times \Omega_{2})$ over a quantum system $\rho \in Z(\HS{H})$.

\subsubsection{Uncertainty Relation for Errors}

Under the assumptions as above, for any pair of localised observables $A, B \in \qob{H}{\rho}$, the relation
\begin{equation}\label{ineq:urel_error_joint}
\err{A}{M}{\rho}\, \err{B}{N}{\rho}
	\geq \sqrt{ \mathcal{R}^{2} + \mathcal{I}^{2} }
\end{equation}
holds with the contributors to the lower bound being
\begin{align}\label{def:urel_joint_real}
\mathcal{R}
	&\defeq \dlev{ \frac{\{A,B\}}{2} }{\!\rho} - \dlinpr{ \pf{M}{\rho} A }{ \pf{M}{\rho} B }{M\!\rho} \notag \\
	&\qquad - \dlinpr{ \pf{N}{\rho} A }{ \pf{N}{\rho} B }{N\!\rho} + \dlinpr{ \pf{M}{\rho} A }{ \pf{N}{\rho} B }{J\!\rho}
\end{align}
and
\begin{align}\label{def:urel_joint_imaginary}
\mathcal{I}
	\defeq \dlev{ \frac{[A,B]}{2i} }{\!\rho} - \dlev{ \frac{[\pb{M}{\rho}\pf{M}{\rho}A,B]}{2i} }{\!\rho} - \dlev{ \frac{[A,\pb{N}{\rho}\pf{N}{\rho}B]}{2i} }{\!\rho}.
\end{align}
Here, the abbreviated notation
\begin{equation}\label{abbr:correlation_joint}
\dlinpr{ \pf{M}{\rho} A }{ \pf{N}{\rho} B }{J\!\rho}
	\defeq \dlinpr{ \spb{\pi_{1}} \pf{M}{\rho} A }{ \spb{\pi_{2}} \pf{N}{\rho} B }{J\!\rho}
\end{equation}
is introduced on the ground of the identifications of the classical observables $f \simeq \spb{\pi_{1}}f$ and $g \simeq \spb{\pi_{2}}g$ mentioned earlier.

\subsubsection{Uncertainty Relation for Errors under Local Representability}

Under the same settings as in the previous relation \eqref{ineq:urel_error_joint}, assume moreover that the observables $A \in \ran{\pb{M}{\rho}}$ and $B \in \ran{\pb{N}{\rho}}$ are both locally representable with respect to the measurements $M$ and $N$ over $\rho$.  Then, the inequality
\begin{equation}\label{ineq:urel_error_representability_joint}
\rerr{A}{M}{\rho}\, \rerr{B}{N}{\rho}
	\geq \sqrt{ \tilde{\mathcal{R}}^{2} + \mathcal{I}_{0}^{2} }
\end{equation}
holds with the contributors to the lower bound being
\begin{equation}\label{def:urel_representability_joint_real}
\tilde{\mathcal{R}}
	\defeq \dlev{ \frac{\{A,B\}}{2} }{\!\rho} - \dlinpr{ \invpb{M}{\rho}A }{ \invpb{N}{\rho}B }{J\!\rho}
\end{equation}
and
\begin{equation}\label{def:urel_imaginary_representability}
\mathcal{I}_{0}
	\defeq \dlev{ \frac{[A,B]}{2i} }{\!\!\rho}.
\end{equation}
Here, the notation
\begin{align}
\dlinpr{ \invpb{M}{\rho}A }{ \invpb{N}{\rho}B }{J\!\rho}
	\defeq& \dlinpr{ \spb{\pi_{1}} \invpb{M}{\rho} A }{ \spb{\pi_{2}} \invpb{N}{\rho} B }{J\!\rho} \notag \\
	=& \dlinpr{ \invpb{J}{\rho} A }{ \invpb{J}{\rho} B }{J\!\rho}
\end{align}
is introduced on the ground of the identifications mentioned earlier, where the second equality follows from the condition \ref{char:partial-inverse_composition_local_3} in Sec.~\ref{sec:partial-inverse:composition}, given the fact that the pullbacks of the projections $\pi_{i}$, $i = 1, 2$, are isometries

\subsubsection{Relation for the Gauge}

The proofs of the relations \eqref{ineq:urel_error_joint} and \eqref{ineq:urel_error_representability_joint} may essentially be understood as mere corollaries to the celebrated Cauchy--Schwarz inequality.  More explicitly, in view of the fact that the gauge \eqref{eq:decomposition_gauge_representability}, which may itself be interpreted as another definition of error in which the classical observable is yet to be optimised, admits a description $\varepsilon_{\rho}(A,f;M) =  \inpr{ (X_{(A,f)}, f) }{ (X_{(A,f)}, f) }^{1/2}$ in terms of the semi-inner product (\textit{i.e.}, positive semi-definite symmetric sesquilinear form)
\begin{equation}\label{def:aux_semi-inner_product}
\inpr{ (X, f) }{ (Y, g) }
	\defeq \lev{ X^{\dagger}Y }{\rho} + \lev{ f^{\dagger} g }{M\!\rho} - \lev{ M^{\prime}f^{\dagger} M^{\prime}g }{\rho}
\end{equation}
defined for the products of Hilbert-space operators $X$ and complex functions $f$ with the shorthand $X_{(A,f)} \defeq A - \pb{M}{\rho}f$, the Cauchy--Schwarz inequality applied to \eqref{def:aux_semi-inner_product} entails
\begin{equation}\label{ineq:gauge}
\varepsilon_{\rho}(A,f;M)\, \varepsilon_{\rho}(B,g;N)
	\geq \sqrt{ R^{2} + I^{2} }
\end{equation}
with
\begin{align}
R
	\defeq \dlev{ \frac{\{A,B\}}{2} }{\!\rho} &- \dlinpr{ f }{ \pf{M}{\rho} B }{M\!\rho} \notag \\
	&- \dlinpr{ \pf{N}{\rho}A }{ g }{N\!\rho} + \dlinpr{ f }{ g }{J\!\rho}
\end{align}
and
\begin{align}
I
	\defeq \dlev{ \frac{[A,B]}{2i} }{\!\rho} - \dlev{ \frac{[\pb{M}{\rho}f,B]}{2i} }{\!\rho} - \dlev{ \frac{[A,\pb{N}{\rho}g]}{2i} }{\!\rho}
\end{align}
being the contributors to the lower bound, where the abbreviated notation
\begin{equation}
\dlinpr{ f }{ g }{J\!\rho}
	\defeq \dlinpr{ \spb{\pi_{1}} f }{ \spb{\pi_{2}} g }{J\!\rho}
\end{equation}
is introduced on the ground of the identifications mentioned earlier;  the two relations \eqref{ineq:urel_error_joint} and \eqref{ineq:urel_error_representability_joint} then follow from the above inequality \eqref{ineq:gauge} with the respective choices of the optimal classical observables, \textit{i.e.}, the choice $f = \pf{M}{\rho}A$ and $g = \pf{N}{\rho}B$ dictated by the pushforward for the former, whereas the choice $f = \invpb{M}{\rho}A$ and $g = \invpb{N}{\rho}B$ dictated by the partial inverse of the pullback for the latter.

\subsubsection{Uncertainty Relation for Local Representatives}

The relations \eqref{ineq:urel_error_joint} and \eqref{ineq:urel_error_representability_joint} entail as a corollary another noteworthy relation regarding the statistical cost of reconstruction of quantum observables.

Under the same assumptions as in \eqref{ineq:urel_error_representability_joint}, namely, provided that the observables $A \in \ran{\pb{M}{\rho}}$ and $B \in \ran{\pb{N}{\rho}}$ are both locally representable by the respective measurements $M$ and $N$, the inequality
\begin{equation}\label{ineq:uncert_joint_representability_2}
\lstdv{f}{M\!\rho}\, \lstdv{g}{N\!\rho}
	\geq \sqrt{ \left( \abs{\tilde{\mathcal{R}}} + \abs{\mathcal{R}_{0}} \right)^{2} + 4\,\mathcal{I}_{0}^{2} }
\end{equation}
holds for any pair of $\rho$-representatives $f$ of $A = \pb{M}{\rho}f$ and $g$ of $B = \pb{N}{\rho}g$, where the contributors $\tilde{\mathcal{R}}$ and $\mathcal{I}_{0}$ to the lower bound are respectively given by \eqref{def:urel_representability_joint_real} and \eqref{def:urel_imaginary_representability}, and the contributor
\begin{equation}\label{def:quantum-covariance}
\mathcal{R}_{0}
	\defeq \dlev{ \frac{\{A,B\}}{2} }{\!\!\rho} - \lev{A}{\rho} \lev{B}{\rho}
\end{equation}
is the familiar quantum covariance defined for the pair of quantum observables concerned.

A simple proof would be to reiteratively apply the Cauchy--Schwarz inequality to the right-hand side of the inequality \eqref{ev:standard-deviation_lower_bound}, thereby revealing
\begin{align}
&\lstdv{f}{M\!\rho}^{2}\, \lstdv{g}{N\!\rho}^{2} \notag \\
	&\quad \geq \dabs{ \rerr{A}{M}{\rho}\,\rerr{B}{N}{\rho} + \lstdv{A}{\rho}\,\lstdv{B}{\rho} }^{2} \notag \\
	&\quad \geq \left( \tilde{\mathcal{R}}^{2} + \mathcal{I}_{0}^{2} \right) + 2 \sqrt{ \tilde{\mathcal{R}}^{2} + \mathcal{I}_{0}^{2} } \sqrt{ \mathcal{R}_{0}^{2} + \mathcal{I}_{0}^{2} }  + \left( \mathcal{R}_{0}^{2} + \mathcal{I}_{0}^{2} \right) \notag \\
	&\quad \geq \left( \tilde{\mathcal{R}}^{2} + \mathcal{I}_{0}^{2} \right) + 2 \left(\abs{ \tilde{\mathcal{R}} } \abs{ \mathcal{R}_{0} } + \mathcal{I}_{0}^{2} \right)  + \left( \mathcal{R}_{0}^{2} + \mathcal{I}_{0}^{2} \right) \notag \\
	&\quad = \left( \abs{\tilde{\mathcal{R}}} + \abs{\mathcal{R}_{0}} \right)^{2} + 4\,\mathcal{I}_{0}^{2},
\end{align}
where the second inequality is due to the Cauchy--Schwarz inequality applied to both the relations \eqref{ineq:urel_error_representability_joint} and the relation
\begin{equation}\label{ineq:urel_Schroedinger}
\lstdv{A}{\rho}\, \lstdv{B}{\rho} \geq \sqrt{\mathcal{R}_{0}^2 + \mathcal{I}_{0}^2},
\end{equation}
known as the Schr{\"o}dinger relation \cite{Schroedinger_1930}, which is in fact a special case of the relation \eqref{ineq:urel_error_joint} when the measurements $M$ and $N$ are trivial (see Sec.~\ref{sec:quantum-indeterminacy}).

It goes without saying that the minimum of the left-hand side of the uncertainty relation \eqref{ineq:uncert_joint_representability_2} is uniquely attained by the optimal choice of the $\rho$-representative, \textit{i.e.}, the partial inverses $f = \invpb{M}{\rho}A$ and $g = \invpb{N}{\rho}B$ of the observables concerned.

\section{Uncertainty Relations for Error and Disturbance\label{sec:urel_error-disturbance}}

The relations for error--disturbance and that for sequential measurements, which are the main results presented in this paper, may be understood as corollaries to the relations \eqref{ineq:urel_error_joint}, \eqref{ineq:urel_error_representability_joint}, and \eqref{ineq:uncert_joint_representability_2} regarding a pair of quantum measurements admitting a local joint-measurement.

\subsection{Observer Effect}

Let $M : Z(\HS{H}) \to W(\Omega_{1})$ be a quantum measurement and let $\Theta : Z(\HS{H}) \to Z(\mathcal{K})$ denote its observer effect, \textit{i.e.}, the inevitable quantum process on the system $\HS{H}$ that the measurement $M$ induces.  It is to be reemphasised that the quantum process $\Theta$ need not be confined to those that end in the same quantum system $\HS{K} = \HS{H}$ as the initial system $\HS{H}$;  the process may result in quantum states pertaining to different systems $\HS{K} \neq \HS{H}$ (such is the case for many situations in physics, \textit{e.g.}, measurement through high energy collision involving particle decays).  Note that these kinds of general treatment were not available for some of the previous formulations of the uncertainty relation regarding observer effects (including the formulation of Ozawa \cite{Ozawa_2003}).

\subsubsection{Sequential Measurement}

Consider a sequential measurement in which the primary measurement is followed by a secondary measurement;  more explicitly, the secondary measurement $L$ is performed after the system underwent the process $\Theta : Z(\HS{H}) \to Z(\mathcal{K})$ caused by the primary measurement $M : Z(\HS{H}) \to W(\Omega_{1})$.  As such, the description of the corresponding map is given by $L : Z(\mathcal{K}) \to W(\Omega_{2})$ where $W(\Omega_{2})$ is the space of probability distributions on the outcomes of the secondary measurement.

\begin{figure}
\includegraphics[hiresbb,clip,width=0.40\textwidth,keepaspectratio,pagebox=artbox]{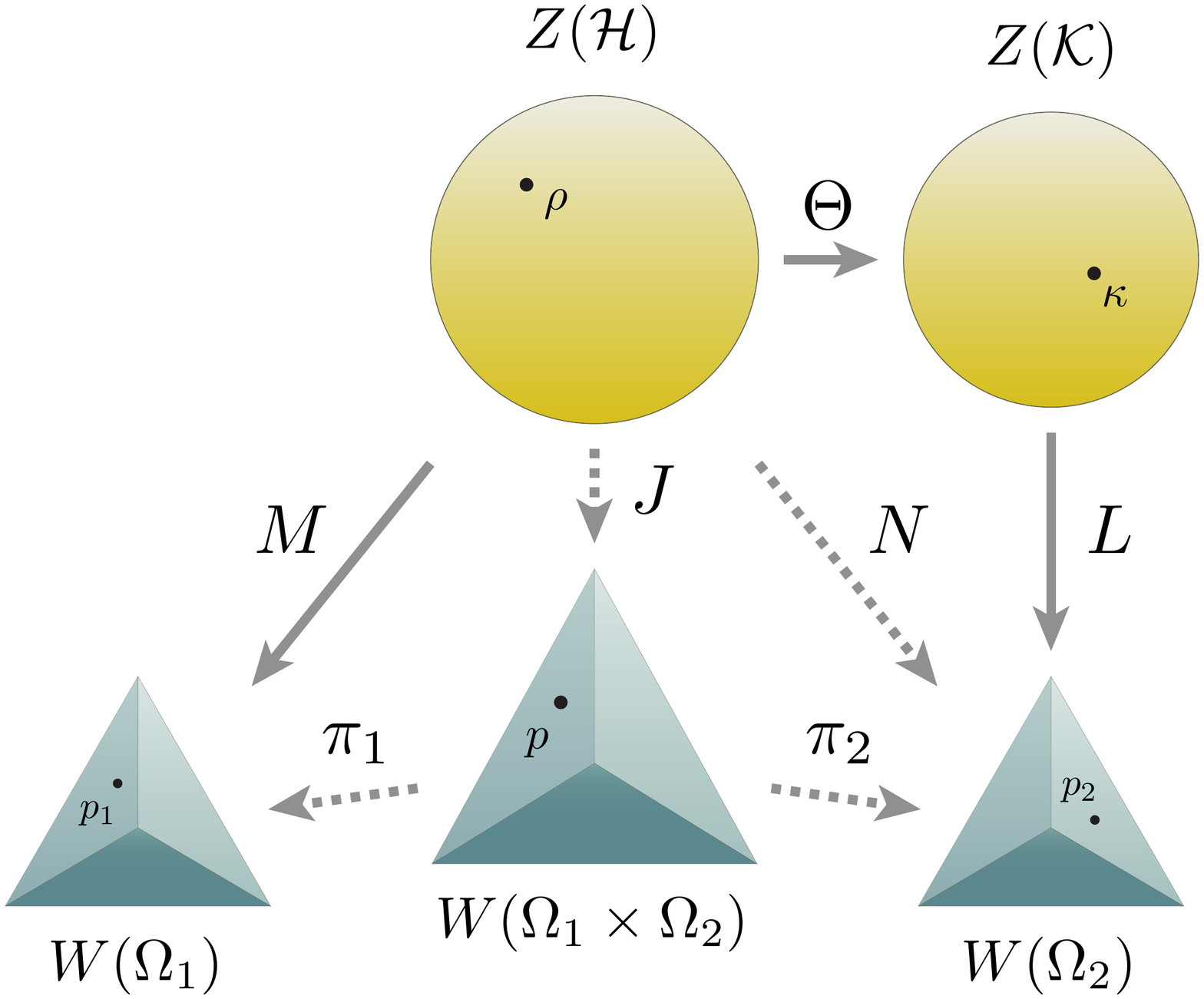}%
\caption{The operational structure of the relation for error and disturbance in terms of sequential measurement.  The primary measurement $M$ on the quantum state $\rho \in Z(\HS{H})$, which results in the probability distribution over the sample space $p_{1} \defeq M\!\rho \in W(\Omega_{1})$, inevitably causes a change $\Theta$ of the initial state into $\kappa \defeq \Theta\hspace{-0.05em}\rho \in Z(\mathcal{K})$ (observer effect).  The secondary measurement $L$ performed over the resultant state $\Theta\hspace{-0.05em}\rho$ yields a probability distribution $p_{2} \defeq L \kappa = (L \circ \Theta)\rho \in W(\Omega_{2})$.  This procedure of sequential measurement operationally entails a joint distribution in $W(\Omega_{1} \times \Omega_{2})$ associated with the joint measurement of the primary $M$ and the secondary $N = L \circ \Theta$ composite measurement.}
\label{fig:sequential-measurement}
\end{figure}

Viewed from the original state space  $Z(\HS{H})$, the procedure of the sequential measurement $M$ followed by $L$ operationally entails a joint measurement $J : Z(\HS{H}) \to W(\Omega_{1} \times \Omega_{2})$ describing the joint behaviour of the primary $M = \pi_{1} \circ J$ and the secondary  $N \defeq L \circ \Theta = \pi_{2} \circ J$ composite measurements, where $\pi_{i}$, $i = 1, 2$, are respectively the projections \eqref{def:projection_1} and \eqref{def:projection_2} introduced earlier (see FIG.~\ref{fig:sequential-measurement}).  Note that the quantum process $\Theta$ describing the observer effect must be constrained strictly by the measurement $M$ in such a manner that the measurement $M$ together with the composite measurement $L \circ \Theta$ must entail a joint measurement $J$ describing the joint behaviour of both the outcomes for arbitrary choices of the secondary measurement $L$.

\subsubsection{Disturbances}

At this point, the characterisation \eqref{char:disturbance} of the disturbance dictates that one may choose a sequence of secondary measurements $L_{n}$ in such a way that the error of the composite measurement converges to the infimum
\begin{equation}\label{eq:disturbance_sequential-measurement}
\lim_{n\to\infty}\err{B}{L_{n} \circ \Theta}{\rho} = \dst{B}{\Theta}{\rho}
\end{equation}
so that the limit coincides with the disturbance of the observable $B$ caused by the process $\Theta$.
In a parallel fashion, the characterisation \eqref{char:disturbance_representability} of the disturbance under local representability entails the existence of a sequence of secondary measurements $L_{n}$ regarding which the disturbance for local representability
\begin{equation}\label{eq:disturbance_sequential-measurement_representability}
\lim_{n\to\infty} \rerr{B}{L_{n} \circ \Theta}{\rho} = \rdst{B}{\Theta}{\rho}
\end{equation}
is described by the limit of errors of the composite measurement.

\subsection{Uncertainty Relation for Error and Disturbance}

From the above observations, one immediately obtains the uncertainty relation for error and disturbance
\begin{equation}\label{ineq:urel_error-disturbance}
\err{A}{M}{\rho}\, \dst{B}{\Theta}{\rho}
	\geq \sqrt{ \mathcal{R}^{2} + \mathcal{I}^{2} }
\end{equation}
as a corollary to the relation \eqref{ineq:urel_error_joint} by simply replacing the measurement $N_{n} = L_{n} \circ \Theta$ with the sequence of composite measurements with $L_{n}$ satisfying \eqref{eq:disturbance_sequential-measurement} and taking the limit;  more explicitly, the contributors to the lower bound reads
\begin{align}\label{def:urel_error-disturbance_real}
\mathcal{R}
	&\defeq \dlev{ \frac{\{A,B\}}{2} }{\!\rho} - \dlinpr{ \pf{M}{\rho} A }{ \pf{M}{\rho} B }{M\!\rho} \notag \\
	&\qquad - \dlinpr{ \pf{\Theta}{\rho} A }{ \pf{\Theta}{\rho} B }{\Theta\hspace{-0.05em}\rho} + \dlinpr{ \pf{M}{\rho} A }{ \pf{N}{\rho} B }{J\!\rho}
\end{align}
and
\begin{align}\label{def:urel_error-disturbance_imaginary}
\mathcal{I}
	\defeq \dlev{ \frac{[A,B]}{2i} }{\!\rho} - \dlev{ \frac{[\dpb{M}{\rho}\dpf{M}{\rho}A,B]}{2i} }{\!\rho} - \dlev{ \frac{[A,\dpb{\Theta}{\rho}\dpf{\Theta}{\rho}B]}{2i} }{\!\rho}
\end{align}
with the abbreviated notation \eqref{abbr:correlation_joint} introduced earlier.  Here, the third terms of the right-hand sides of \eqref{def:urel_joint_real} and \eqref{def:urel_joint_imaginary} are due to the relation $\spb{(N_{n})}\spf{(N_{n})} B = \spb{(L_{n} \circ \Theta)} \spf{(L_{n} \circ \Theta)} B = \spb{\Theta} \spb{(L_{n})} \spf{(L_{n})}\spf{\Theta} B \to \spb{\Theta}\spf{\Theta}B$ as $n \to \infty$, in which the composition laws of the pullback \eqref{eq:composition_pullback} and the pushforward \eqref{eq:composition_pushforward} is employed in the second equality, and the convergence follows from the continuity of the pullback (and pushforward), noting the construction of $L_{n}$ and the condition  \ref{char:errorless-measurement_3} of the lossless process discussed in Sec.~\ref{sec:loss:lossless:measurement}.

Specifically, the uncertainty relation \eqref{ineq:urel_error-disturbance} entails as a trivial corollary the simplified form
\begin{equation}\label{ineq:urel_error-disturbance_simple}
\err{A}{M}{\rho}\, \dst{B}{\Theta}{\rho}
	\geq \abs{ \mathcal{I} },
\end{equation}
in which the description of the lower bound becomes independent of the choice of the sequence of secondary measurements $L_{n}$ as expected from the definitions of the error and disturbance.

\subsection{Uncertainty Relation for Error and Disturbance under Local Representability\label{sec:urel_error-disturbance:representability}}

Under the same settings as in the previous relation \eqref{ineq:urel_error-disturbance}, assume moreover that the observables $A \in \ran{\pb{M}{\rho}}$ and $B \in \ran{\pb{\Theta}{\rho}}$ are both locally representable with respect to the measurements $M$ and the resultant observer effect $\Theta$ over $\rho$.  One then finds, form parallel logic as above, the uncertainty relation for error and disturbance
\begin{equation}\label{ineq:urel_error-disturbance_representability}
\rerr{A}{M}{\rho}\, \rdst{B}{\Theta}{\rho}
	\geq \sqrt{ \tilde{\mathcal{R}}^{2} + \mathcal{I}_{0}^{2} }
\end{equation}
under local representability by simply replacing the measurement $N_{n} = L_{n} \circ \Theta$ with the composite measurement defined for the choice of a sequence of secondary measurements $L_{n}$ satisfying \eqref{eq:disturbance_sequential-measurement_representability} in the relation \eqref{ineq:urel_error_representability_joint}.  Here, the contributors $\tilde{\mathcal{R}}$ and $\mathcal{I}_{0}$ to the lower bound are respectively given by \eqref{def:urel_representability_joint_real} and \eqref{def:urel_imaginary_representability} as before.

As above, the uncertainty relation \eqref{ineq:urel_error-disturbance_representability} specifically entails as a trivial corollary the simplified form
\begin{equation}\label{ineq:urel_error-disturbance_representability_simple}
\rerr{A}{M}{\rho}\, \rdst{B}{\Theta}{\rho}
	\geq \abs{ \mathcal{I}_{0} },
\end{equation}
in which the description of the lower bound becomes independent of the choice of the sequence of secondary measurements $L_{n}$ as expected from the definitions of the error and disturbance.

\subsection{Uncertainty Relation for Costs in Sequential Measurement under Local Representability}

As has been discussed so far, the sequential measurement of a primary measurement $M$ followed by a secondary measurement $L$ operationally entails a (local) joint measurement $J$ of the measurement $M$ and the composite measurement $N = L \circ \Theta$, where $\Theta$ is the quantum process induced by the act of the primary measurement.  It goes without saying that the relation \eqref{ineq:uncert_joint_representability_2} for local representatives may thus be applied to the settings of sequential measurement as a special case, in which the classical observable $f$ is meant to reconstruct the quantum observable $A$ through the primary measurement $M$ and $g$ is meant to reconstruct $B$ through the secondary measurement $N = L \circ \Theta$.

\section{The Uncertainty Principle: A No-Go Theorem\label{sec:uncertainty-principle}}

The uncertainty relation \eqref{ineq:urel_error_joint} for error and that \eqref{ineq:urel_error-disturbance} for error--disturbance connote a potential violation of the na{\"i}ve lower bound $\abs{ \lev{ [A, B] }{\rho} }/2$ for certain choices of quantum measurements and the induced observer effects; indeed, an errorless measurement of either observable, which is always attainable by the projection measurement associated with it, trivially violates it.   Another immediate way to observe this would be to apply the fact that the relations \eqref{ineq:urel_error_joint} and \eqref{ineq:urel_error-disturbance} are respectively tighter than Ozawa's relations for joint POVM measurements and error--disturbance (as will be discussed shortly);  since the latter relations are known to violate the bound, the former relations do so whenever this is the case.

Still, Heisenberg's philosophy of the uncertainty principle prevails, albeit perhaps in a weaker sense than was originally intended, as one shall see in the presentation below.

\subsection{The Uncertainty Principle for Quantum Measurements}

The uncertainty relation \eqref{ineq:urel_error_joint} for errors entails a no-go theorem, stating that whenever $\lev{ [A, B] }{\rho} \neq 0$ is non-vanishing for a pair of quantum observables $A$ and $B$, their local joint errorless measurement is impossible over the said state.  This may be observed at once by resorting to argumentum ad absurdum;  if there were such a pair of measurements admitting a joint local description, then the relation \eqref{ineq:urel_error_joint} combined with the condition \ref{char:errorless-measurement_3} for errorless measurements in Sec.~\ref{sec:loss:lossless:measurement} would lead to a contradiction $0 \geq \sqrt{\abs{ 0 }^{2} + \abs{ \lev{ [A, B] }{\rho}/2i }^{2}} > 0$.  A simple corollary to this is that, for non-trivial (\textit{i.e.}, $\mathrm{dim}(\HS{H}) \geq 2$) quantum systems, there exists no measurement that is capable of measuring every observable errorlessly over every state singlehandedly.

Given the equivalence of the errorless condition between the two definitions \eqref{def:error_quantum-measurement} and \eqref{def:error_quantum-measurement_representability} of the error, the uncertainty principle in the (weaker) sense of this paper regarding the former error is valid verbatim for the latter error for local representability as well.  Meanwhile, the relation \eqref{ineq:urel_error_representability_joint} under local representability differs from the relation \eqref{ineq:urel_error_joint} in that it always respects the na{\"i}ve lower bound:  as long as the pair of observables $A$ and $B$ are locally representable by a pair of measurements that admit a local joint measurement, neither of the errors may vanish whenever $\lev{ [A, B] }{\rho} \neq 0$ is non-vanishing (as opposed to the relation \eqref{ineq:urel_error_joint}, in which either, but not both, of the errors may does so).

\subsection{The Uncertainty Principle for Observer Effects}

In a parallel manner, the uncertainty relation \eqref{ineq:urel_error-disturbance} for error--disturbance entails a no-go theorem, stating that whenever $\lev{ [A, B] }{\rho} \neq 0$ is non-vanishing for a pair of quantum observables $A$ and $B$, errorless measurement of either of the observable is impossible without causing disturbance to the other over the said state.  Indeed, reductio ad absurdum states:  if there were such a measurement, the relation \eqref{ineq:urel_error-disturbance} combined with the condition \ref{char:errorless-measurement_3} for errorless measurements in Sec.~\ref{sec:loss:lossless:measurement} and the parallel condition \ref{char:nondisturbing-process_3} for non-disturbing processes in Sec.~\ref{sec:loss:lossless:process} would lead to a contradiction $0 \geq \sqrt{\abs{ 0 }^{2} + \abs{ \lev{ [A, B] }{\rho}/2i }^{2}} > 0$.  A simple corollary to this is that, for non-trivial (\textit{i.e.}, $\mathrm{dim}(\HS{H}) \geq 2$) quantum systems, there exists no quantum measurement without observer effect.

As previous, given the equivalence of the non-disturbing condition between the two definitions \eqref{def:disturbance_quantum-process} and \eqref{def:disturbance_quantum-process_representability} of the disturbance, the uncertainty principle in the (weaker) sense of this paper regarding the former definitions of error and disturbance is valid verbatim for the latter definitions of error and disturbance for local representability as well.  Meanwhile, the relation \eqref{ineq:urel_error-disturbance_representability} under local representability differs from the relation \eqref{ineq:urel_error-disturbance} in that it always respects the na{\"i}ve lower bound:  as long as the pair of observables $A$ and $B$ are locally representable by the measurement and its resultant process, neither of the error and disturbance may vanish whenever $\lev{ [A, B] }{\rho} \neq 0$ is non-vanishing (as opposed to the relation \eqref{ineq:urel_error-disturbance}, in which either, but not both, of the error or disturbance may does so).

\section{Reference to Other Relations\label{sec:reference_to_other_relations}}

Below, some of the notable previously known uncertainty relations are examined in the new light of the relations presented in this paper.

\subsection{Quantum Indeterminacy\label{sec:quantum-indeterminacy}}

The celebrated Kennard--Robertson relation \eqref{ineq:urel_Kennard-Robertson}, which is known to have very little to do with the concept of measurement, in fact emerges as a trivial case of the relation \eqref{ineq:urel_error_joint} for errors.  A quantum measurement $M$ may be called trivial, or non-informative, when it is a constant map, \textit{i.e.}, there exists a fixed probability distribution $p_{0} \in W(\Omega)$ such that $M\!\rho = p_{0}$ holds for all quantum states $\rho \in Z(\HS{H})$.  One may readily verify that the pullback and pushforward of a trivial measurement $M$ are characterised by the identity operator $\dpb{M}{\rho}f = \lev{f}{M\!\rho}$ and the constant function $\pf{M}{\rho}A = \lev{A}{\rho}$, each weighted by the expectation values of the observables concerned.  Triviality of measurement thus reduces the error \eqref{def:error_quantum-measurement} to the standard deviation $\err{A}{M}{\rho} = \lstdv{A}{\rho}$, further bringing the overall relation \eqref{ineq:urel_error_joint} to the Schr{\"o}dinger relation \eqref{ineq:urel_Schroedinger}, to which the Kennard--Robertson relation \eqref{ineq:urel_Kennard-Robertson} is yet another trivial corollary.

\subsection{The Ozawa Relations}

Other notable corollaries to the special cases of the general relations \eqref{ineq:urel_error_joint} and \eqref{ineq:urel_error-disturbance} are respectively the Ozawa relation \cite{Ozawa_2004_01} regarding errors of joint measurements (as well as its recent modification \cite{Ozawa_2019}) and his relation \cite{Ozawa_2003} for error--disturbance.  For this, it should be noted that Ozawa's error $\varepsilon$ is defined for the special class of quantum measurements that yield real outcomes $\Omega = \mathbb{R}$, and that the error \eqref{def:error_quantum-measurement} and disturbance \eqref{def:disturbance_quantum-process} are mathematically well-defined whenever Ozawa's error and disturbance are;  in fact, the error \eqref{def:error_quantum-measurement} and disturbance \eqref{def:disturbance_quantum-process} are respectively never greater than the counterparts in Ozawa's definition.

\subsubsection{The Ozawa Relation for Errors of Joint Measurement}

Given the fact that the (global) joint measurability, upon which many of the previous formulations (including Ozawa's) are founded, is a stronger condition than that of the local joint measurability introduced in this paper, let $J$ be a (global) joint POVM measurement with outcomes in the product space $\Omega = \mathbb{R}^{2}$.  Provided that Ozawa's errors $\epsilon(A)$ and $\epsilon(B)$ are both well-defined over the state $\rho$ for the marginal POVM measurements $M_{1}$ and $M_{2}$ of $J$, respectively, one finds the chain of inequalities $\epsilon(A)\, \epsilon(B) \geq \err{A}{M_{1}}{\rho}\, \err{B}{M_{2}}{\rho} \geq ( \mathcal{R}^2 + \mathcal{I}^2 )^{1/2} \geq \abs{ \mathcal{I} } \geq \abs{ \lev{ [A,B] }{\rho} } /2 - \epsilon(A)\,\lstdv{B}{\rho} - \lstdv{A}{\rho}\,\epsilon(B)$ with $\mathcal{R}$ and $\mathcal{I}$ being respectively the semiclassical \eqref{def:urel_joint_real} and quantum \eqref{def:urel_joint_imaginary} contributions to the lower bound of the product of the errors.  Here, note that the left-most- and the right-most-hand sides of the above chain is equivalent to the Ozawa relation, whereas the inequality in the middle is the relation \eqref{ineq:urel_error_joint} under local joint-measurability.

\subsubsection{The Ozawa Relation for Error--Disturbance}

With the above remarks in mind, let $M$ be a POVM measurement over $\HS{H}$ with real outcomes $\Omega = \mathbb{R}$, and let $\Theta$ denote the quantum process induced by $M$ that results in the same sate-space $\HS{K} = \HS{H}$ as the original system over which the measurement is performed.  Provided that Ozawa's error $\epsilon(A)$ of the measurement $M$ of an observable $A$ and disturbance $\eta(B)$ of the observable $B$ caused by $\Theta$ are both well-defined over the state $\rho$, one finds the chain of inequalities $\epsilon(A)\, \eta(B) \geq \err{A}{M}{\rho}\, \dst{B}{\Theta}{\rho} \geq ( \mathcal{R}^2 + \mathcal{I}^2 )^{1/2} \geq \abs{ \mathcal{I} } \geq \abs{ \lev{ [A,B] }{\rho} } /2 - \epsilon(A)\,\lstdv{B}{\rho} - \lstdv{A}{\rho}\,\eta(B)$ with $\mathcal{R}$ and $\mathcal{I}$ being respectively the semiclassical \eqref{def:urel_error-disturbance_real} and quantum \eqref{def:urel_error-disturbance_imaginary} contributions to the lower bound of the product of the error and disturbance.  Here, note that the left-most- and the right-most-hand sides of the above chain is equivalent to the Ozawa relation, whereas the inequality in the middle is the relation \eqref{ineq:urel_error-disturbance}.

\subsection{The Arthurs--Kelly--Goodman Relations}

The notion of local representability of observables by a process introduced in this paper can in essence be understood as a more general and universal concept than those of `unbiasedness' found in the operator-theoretic formulation of Arthurs--Kelly--Goodman, or those of `local unbiasedness' found in the framework of estimation theory adopted by Watanabe, Sagawa, and Ueda in formulating their relations.

In this regard, the Arthurs--Kelly--Goodman relations \cite{Arthurs_1965,Arthurs_1988} are other noteworthy instances of the corollaries to the special cases of the universal framework of the author and its resultant relations presented in this paper.  For this, it should be noted that, in view of the fact that Ozawa's definition of error is an extension of AKG's, the latter of which is only relevant when the measurement satisfies the (global) `unbiasedness' condition regarding the measurement of the observables concerned, the same remarks regarding Ozawa's formulation also apply to AKG's.  

\subsubsection{The Arthurs--Kelly--Goodman Relation for Errors}

Given the fact that the concepts of (global) unbiasedness and (global) joint-measurability adopted by AKG are respectively much stronger conditions than that of local representability and local joint-measurability introduced in this paper, let $J$ be a (global) joint measurement with outcomes in $\Omega = \mathbb{R}^{2}$.  Provided that the marginal POVM measurements $M_{1}$ and $M_{2}$ of $J$ are capable of the (global) unbiased measurement of the respective observables $A$ and $B$, and that AKG's errors $\epsilon(A)$ and $\epsilon(B)$ are both well-defined over the state $\rho$, one finds the chain of inequalities $\epsilon(A)\, \epsilon(B) \geq \rerr{A}{M_{1}}{\rho}\, \rerr{B}{M_{2}}{\rho} \geq ( \tilde{\mathcal{R}}^2 + \mathcal{I}_{0}^2 )^{1/2} \geq \abs{ \lev{ [A,B] }{\rho} } /2$ with $\tilde{\mathcal{R}}$ and $\mathcal{I}_{0}$ being respectively the semiclassical \eqref{def:urel_representability_joint_real} and quantum \eqref{def:urel_imaginary_representability} contributions to the lower bound of the product of the errors.  Here, the left-most- and the right-most-hand sides of the above chain is the AKG relation, whereas the inequality in the middle is the relation \eqref{ineq:urel_error_representability_joint} under local representability.

\subsubsection{The Arthurs--Kelly--Goodman Relation for Standard Deviations}

As for the other AKG's relation regarding the standard deviations of the measurement outcomes, one finds under the same settings as above, the chain of inequalities $\lstdv{M_{1}}{\rho}\,\lstdv{M_{2}}{\rho} \geq \lstdv{\tilde{f}_{A}}{M\!\rho}\,\lstdv{\tilde{f}_{B}}{N\!\rho} \geq ( ( \abs{\tilde{\mathcal{R}}} + \abs{\mathcal{R}_{0}} )^{2} + 4\,\mathcal{I}_{0}^{2} )^{1/2} \geq \abs{ \lev{ [A,B] }{\rho} }$ with $\tilde{f}_{A} \defeq \inv{(\spb{M_{1}})} A$ and $\tilde{f}_{B} \defeq \inv{(\spb{M_{2}})} B$ respectively being the optimal $\rho$-representatives of $A$ and $B$ dictated by the partial inverses of the pullbacks of the measurements concerned.  Here, $\tilde{\mathcal{R}}$, $\mathcal{R}_{0}$, and $\mathcal{I}_{0}$ are respectively given by \eqref{def:urel_representability_joint_real}, \eqref{def:quantum-covariance}, and \eqref{def:urel_imaginary_representability}.  The left-most- and the right-most-hand sides of the above chain is the AKG relation for the standard deviations of the measurement outcomes, whereas the inequality in the middle is the relation \eqref{ineq:uncert_joint_representability_2} for the statistical costs of the local representatives presented in this paper.

\subsection{The Watanabe--Sagawa--Ueda Relations}

The Watanabe--Sagawa--Ueda relations \cite{Watanabe_2011, Watanabe_2011_06, Watanabe_2014} adopts the estimation theory \cite{Holevo_1982} as its grounding framework.  As such, they presume additional structures within their frameworks (most notably, differential structures), which the framework of this paper does not necessitate;  in this regard, the framework of this paper is more universal than theirs (it is also worth noting that their results are only accountable for quantum measurements on finite-dimensional Hilbert spaces, whereas the results of this paper has no such limits).

Specifically, as mentioned above, the notion of local representability (and local representatives) introduced in this paper is found to be a more universal concept than those of local unbiasedness (and locally unbiased estimators) adopted in their framework;  the former concepts of this paper may be defined without any additional structure (in this case, differential structures in particular), whereas the latter concepts of theirs cannot do without them.

In this regard, the Watanabe--Sagawa--Ueda relation \cite{Watanabe_2011} for errors of quantum measurements and the Watanabe--Ueda relation \cite{Watanabe_2011_06} for error--disturbance may respectively in essence be understood to be corollaries to the special cases of the uncertainty relations \eqref{ineq:urel_error_representability_joint} for errors and that \eqref{ineq:urel_error-disturbance_representability} for error--disturbance under local representability.  Since a detailed discussion on this topic requires much exposure to the estimation theory (especially differential geometry), which is way beyond the scope of this paper, only an outline of the results will be given below;  a comprehensive description will be given elsewhere in later publications of the author.

\subsubsection{The Watanabe--Sagawa--Ueda Relation for Errors}

The universal formulation of the author may be furnished with additional mathematical structures (differential structures, in particular) so that the standard estimation-theoretic interpretations of the various concepts and results presented so far become available.

Let $\mathbf{M}_{1}$ and $\mathbf{M}_{2}$ be a pair of POVM measurements performed on a quantum system described by a finite-dimensional Hilbert space $\HS{H}$.  With some mathematical assumptions that guarantee the well-definedness and good behaviours of the objects concerned, the details of which will be given elsewhere, assume moreover that the pair $\mathbf{M}_{1}$ and $\mathbf{M}_{2}$ admits a local joint-measurement---which is a natural but non-trivial extension of the standard notion of (global) joint measurements (see Sec.~\ref{sec:urel_error:local-joint-measurability})---by a mediating POVM measurement $\mathbf{M}$ over the quantum state $\rho_{\boldsymbol{\theta}} \in Z(\HS{H})$ with $\boldsymbol{\theta}$ being a parameter of some (multi-dimensional) chart of a given atlas of the quantum-state space $Z(\HS{H})$.

Then, with due consideration to the local representability (or, local unbiasedness in this context) of a pair of observables $\hat{A}$ and $\hat{B}$ regarding the respective measurements $\mathbf{M}_{1}$ and $\mathbf{M}_{2}$, the relation \eqref{ineq:urel_error_representability_joint} for errors under local representability entails, in essence as a corollary to its special case, the relation $\varepsilon(\hat{A};\mathbf{M}_{1})\, \varepsilon(\hat{B};\mathbf{M}_{2}) \geq \tilde{\mathcal{R}}^2 + \mathcal{I}_{0}^2$, granted the local joint-measurability of $\mathbf{M}_{1}$ and $\mathbf{M}_{2}$ mentioned above.  Here, the symbols $\varepsilon(\hat{A};\mathbf{M}_{1})$ and $\varepsilon(\hat{B};\mathbf{M}_{2})$ respectively denote the (refined definitions of the) WSU error of $\mathbf{M}_{1}$ in measuring $\hat{A}$ and that of $\mathbf{M}_{2}$ in measuring $\hat{B}$.  Conforming to their notations, the semiclassical contributor to the lower bound essentially reads $\tilde{\mathcal{R}} \defeq \mathcal{R}_{0} - \mathbf{a} \cdot \inv{J(\mathbf{M})} \mathbf{b}$, where $\mathcal{R}_{0}$ is as in \eqref{def:quantum-covariance} and $J(\mathbf{M})$ is the Fisher information matrix at the parameter $\boldsymbol{\theta}$ associated with the locally mediating POVM measurement $\mathbf{M}$ of the pair of measurements $\mathbf{M}_{1}$ and $\mathbf{M}_{2}$, where the vectors concerned are $\mathbf{a} \defeq \partial_{\boldsymbol{\theta}} \ev{\hat{A}}$ and $\mathbf{b} \defeq \partial_{\boldsymbol{\theta}} \ev{\hat{B}}$ with the dot $\mathbf{x} \cdot \mathbf{y} \defeq \mathbf{x}^{\mathrm{T}} \mathbf{y}$ being the usual vector inner-product.  Its quantum counterpart $\mathcal{I}_{0}$ is as in \eqref{def:urel_imaginary_representability}.  The WSU relation for errors \cite{Watanabe_2011} thus straightforwardly follows, in essence, as a trivial corollary to the above relation by taking $\mathbf{M}_{1} = \mathbf{M}_{2}$ and further omitting the semiclassical term $\tilde{\mathcal{R}}$.

\subsubsection{The Watanabe--Ueda Relation for Error--Disturbance}

The act of the measurement of $\mathbf{M}_{1}$ induces a quantum process over the system on which the measurement is performed (observer effect).  Under the same notations as above, and again with due consideration to the local representability (or, local unbiasedness in this context) of a pair of observables $\hat{A}$ and $\hat{B}$ regarding the measurement $\mathbf{M}_{1}$ and its resultant quantum process, respectively, the relation \eqref{ineq:urel_error-disturbance} for error--disturbance under local representability entails, in essence as a corollary to its special case, the relation $\varepsilon(\hat{A};\mathbf{M}_{1})\, \eta(\hat{B};\mathbf{M}_{1}) \geq \tilde{\mathcal{R}}^2 + \mathcal{I}_{0}^2$.  Here, the symbol $\varepsilon(\hat{A};\mathbf{M}_{1})$ is the same as above, whereas the symbol $\eta(\hat{B};\mathbf{M}_{1})$ denotes the (refined definition of the) WSU disturbance of $\hat{B}$ caused by the quantum process induced by the act of the measurement of $\mathbf{M}_{1}$ (a slight abuse of notation is employed here in denoting the measurement and the measurement process with the same symbol).  The lower bound to their product is dictated by the semiclassical contributor $\tilde{\mathcal{R}} \defeq \mathcal{R}_{0} - \mathbf{a} \cdot \inv{J(\mathbf{M})} \mathbf{b}$ with $\mathcal{R}_{0}$ being \eqref{def:quantum-covariance} and $J(\mathbf{M})$ being the Fisher information matrix at the parameter $\boldsymbol{\theta}$ associated with the joint POVM measurement $\mathbf{M}$ operationally defined by the sequential measurement of the initial measurement $\mathbf{M}_{1}$ followed by an auxiliary secondary measurement (see discussions in Sec.~\ref{sec:urel_error-disturbance:representability}).  As previous, the quantum contributor $\mathcal{I}_{0}$ is as in \eqref{def:urel_imaginary_representability}.  The WSU relation for error--disturbance \cite{Watanabe_2011_06} thus straightforwardly follows, in essence, as a trivial corollary to the above relation by omitting the semiclassical term $\tilde{\mathcal{R}}$.

\begin{acknowledgments}
This work was supported by JSPS Grant-in-Aid for Scientific Research (KAKENHI), Grant Numbers JP18K13468, JP22K13970, and JP20H01906.
\end{acknowledgments}

\bibliography{}

\end{document}